\documentclass[11pt,xcolor=dvipsnames,fleqn]{article}
\DeclareMathAlphabet{\scr}{U}{rsfs}{m}{n}

\pdfoutput=1
\usepackage{scalerel,stackengine}
\usepackage{subfigure}
\usepackage{latexsym}
\usepackage{epsfig}
\usepackage[mathscr]{eucal}
\usepackage{amsfonts}
\usepackage{amscd}
\usepackage{multirow}
\usepackage{cite}
\usepackage{array}
\usepackage{amssymb}
\usepackage{amsthm}
\usepackage[outdir=./]{epstopdf}
\usepackage{colordvi}
\usepackage[centertags]{amsmath}
\usepackage{indentfirst}
\usepackage{geometry}
\usepackage{float} 
\usepackage{enumerate}
\usepackage{graphicx}
\usepackage{booktabs}
\usepackage{theorem}
\usepackage[footnotesize]{caption}
\usepackage{soul}
\usepackage{mcite}
\usepackage{slashed}
\usepackage{braket}
\usepackage{color,xcolor}
\usepackage{bbm}

\usepackage[utf8]{inputenc}
\usepackage{fancyvrb}
\usepackage{framed}
\usepackage{xspace}
\usepackage{todonotes}
\usepackage{siunitx}

\usepackage{psfrag,rotating}
\usepackage{epstopdf}
\numberwithin{figure}{section}
\numberwithin{table}{section}
\def\zz{\mathbb{Z}}

\allowdisplaybreaks

\usepackage{cleveref}
\crefname{chapter}{Chapter}{Chapter}
\crefname{section}{Sec.}{Secs.}
\crefname{table}{Tab.}{Tabs.}
\crefname{figure}{Fig.}{Figs.}
\crefname{equation}{Eq.}{Eqs.}
\crefname{appendix}{Appendix\ }{Appendix\ }

\begin{document}
\pdfoutput=1

\title{
    \vspace*{-3.7cm}
    \vspace*{2.7cm}
    \textbf{Full one-loop radiative corrections to $e^+ e^-\to H^+H^-$ in the inert doublet model \\[4mm]}}

\date{}
\author{
Hamza Abouabid$^{1\,}$\footnote{E-mail:
    \texttt{hamza.abouabid@gmail.com}} ,
Abdesslam Arhrib$^{1\,}$\footnote{E-mail:
    \texttt{aarhrib@gmail.com}} ,
Jaouad El Falaki$^{2\,}$\footnote{E-mail: \texttt{jaouad.elfalaki@gmail.com}},
Bin Gong$^{3,4\,}$\footnote{{E-mail: \texttt{twain@ihep.ac.cn}}},
\\
Wenhai Xie$^{3,4\,}$\footnote{E-mail:
    \texttt{xiewh@ihep.ac.cn}},
Qi-Shu Yan$^{4,5\,}$\footnote{E-mail: \texttt{yanqishu@ucas.ac.cn}}\,\,,
\\[5mm]
{\small\it
$^1$Universit\'{e} Abdelmalek Essaadi, FSTT, B. 416, Tangier, Morocco.} \\[3mm]
{\small\it $^2$ LPTHE, Physics Department, Faculty of Sciences, Ibnou Zohr University, P.O.B. 8106 Agadir, Morocco.} \\[3mm]
{\small\it $^3$ Theory Division, Institute of High Energy Physics, Chinese Academy of Sciences, Beijing 100049, China.} \\[3mm]
{\small\it $^4$School of Physics Sciences, University of Chinese Academy of Sciences, Beijing 100049, China.} \\[3mm]
{\small\it $^5$ Center for Future High Energy Physics, Chinese Academy of Sciences,  Beijing 100049, China.} \\[3mm]
}

\maketitle

\begin{abstract}
We compute the full one-loop radiative corrections
for charged scalar pair production $e^{+}e^{-}\to H^{+}H^{-}$ in the inert doublet model. The on-shell renormalization scheme has been used.
We take into account both the weak contributions as well as the soft and hard QED corrections.
We compute both the real emission and the one-loop virtual corrections using the Feynman diagrammatic method.
The resummed cross section is introduced to cure the Coulomb singularity which occurs in the QED corrections. 
We have analyzed the parameter space of the inert 
doublet model in three scenarios after taking into account 
theoretical constraints, the collider experimental bounds, and dark matter search bounds as well. 
It is found that the weak interaction dominates the radiative corrections, and its size is determined by the triple Higgs coupling $\lambda_{h^0 H^+ H^-}$, which is further connected to the mass of the charged scalar. In the scenario where all the constraints are taken into account, we find that
 for $\sqrt{s}=250$ GeV and $\sqrt{s}=500$ GeV, the weak corrections are around $-6\% \sim-5\%$ and $-10\% \sim -3\%$, respectively. While for $\sqrt{s}=1000$ GeV, the weak corrections can reach $-15\% \sim +25\%$. The new feature is that the weak corrections can be positive near the threshold when the charged scalar is heavier than 470 GeV. 
 Six benchmark points for future collider searches have been proposed.
\end{abstract}
\thispagestyle{empty}
\vfill
\newpage
\setcounter{page}{1}
\section{Introduction}
\label{sec:introduction}
The Standard Model (SM) particle spectrum has been completed with the 
discovery of the Higgs boson on July 4, 2012, by the ATLAS and CMS experiments at CERN \cite{Aad:2012tfa,Chatrchyan:2012xdj}. 
Furthermore, this discovery has confirmed  that  the SM of particle physics  is
the underlying theoretical framework valid at least for  energies up to the electroweak (EW) scale. The two collaborations also  carried out several Higgs boson couplings measurements at the Large Hadron Collider (LHC) during run-1 and run-2 such as {the couplings of the Higgs boson to top quarks  \cite{CMS:2018uxb, ATLAS:2018mme}, tau leptons \cite{CMS:2017zyp, ATLAS:2018ynr},
 bottom quarks} \cite{ATLAS:2018kot, CMS:2018nsn}, and all the electroweak gauge bosons, 
including the decays to $ZZ^*$  \cite{CMS:2018gwt, ATLAS:2013dos}, 
$WW^*$ \cite{ATLAS:2018xbv, ATLAS:2019vrd, CMS:2018zzl, CMS:2020dvg}, and $\gamma \gamma$ \cite{ATLAS:2018hxb}.  {In addition}, recently upper limits have been set on the $h^0\rightarrow \gamma Z$ signal strength \cite{ATLAS:2020qcv, CMS:2018myz} and 
on the Higgs boson production cross section times branching fraction to muons \cite{ATLAS:2018kbw, CMS:2018nak}. 

The aforementioned measurements will be improved at future experiments such as 
the High-Luminosity LHC \cite{Cepeda:2019klc, deBlas:2019rxi}, scheduled to operate from 2029,  where the Higgs boson couplings are projected to be improved to a precision level of $5 - 10\%$. In addition, the experimental uncertainties will be further reduced
in the clean environment of the future lepton colliders, such as the  International Linear Collider (ILC) \cite{Moortgat-Picka:2015yla, Bambade:2019fyw}, 
the  Circular  Electron  Positron  Collider (CEPC) \cite{CEPC-SPPCStudyGroup:2015csa}, the Compact Linear Collider (CLIC) \cite{Battaglia:2004mw, Aicheler:2012bya, Linssen:2012hp}, and the Future Circular Collider \cite{Gomez-Ceballos:2013zzn}. For example, at the ILC,
  with a  c.m.energy of  about 250 GeV and a  luminosity of $2\,ab^{-1}$, some Higgs boson couplings will most likely be  measured at a  precision level of  $1\%$  for $h^0\to b\bar{b}$  and below  $1\%$ for  $h^0\rightarrow ZZ, WW$ \cite{Fujii:2017vwa}. 
  
Although new physics beyond the SM has not yet been established by the current LHC dataset, it is necessary in order to understand several puzzles of the SM and the observed Universe, such as what is the nature of dark matter, what is the origin of neutrino masses, how to stabilize the vacuum of the SM, and so on. From the viewpoint of effective theory, the success of the SM can only be understood as the low energy limit of a more fundamental theory. In this more fundamental theory, there must be extra sectors not included by the SM. For example, the Higgs sector of the SM is assumed to be minimal and composed of only one Higgs doublet. In order to solve the problem of $CP$ violation, it is well motivated to extend the Higgs sector of the SM by introducing one more doublet \cite{Lee:1973iz} and the Higgs potential sector can be extended, such a theory model is called the two Higgs doublet model (2HDM). A comprehensive review of the 2HDM can be found in the reference \cite{Branco:2011iw}.

In order to offer a dark matter candidate, a $\zz_2$ symmetry is introduced into the 2HDM, which leads to the so-called inert doublet model (IDM). In the IDM, the second doublet does not develop a vacuum expectation value (VEV) nor does it have a direct coupling to the SM fermions. In such a $\zz_2$ symmetry,  {{the fermions, gauge bosons, and the SM Higgs doublet are invariant, and the second doublet is odd, i.e. $H_2 \to - H_2$}}.  The  $\zz_2$  symmetry ensures that this extra doublet has
no direct coupling to the SM fermions at both tree and loop levels, and its lightest stable neutral component  may play the role of a dark matter candidate.
The IDM was proposed by E. Ma et al. \cite{Deshpande:1977rw} and it was initially suggested for studies on electroweak symmetry breaking. This model is very appealing because
it can generate small neutrino masses \cite{Ma:2006km}, can provide a dark matter candidate \cite{Gustafsson:2007pc,Hambye:2007vf,Agrawal:2008xz,Dolle:2009fn,Cao:2007rm,LopezHonorez:2006gr,Goudelis:2013uca} and can solve the naturalness problem \cite{Barbieri:2006dq}. The phenomenology of the IDM has been extensively studied  in the literature in  the  context of  the LHC and
at future Higgs factories such as the ILC  or CLIC \cite{Dolle:2009ft,Aoki:2013lhm,Arhrib:2012ia, Krawczyk:2013jta,Arhrib:2014pva, Datta:2016nfz,Dutta:2017lny,Kalinowski:2018ylg,Kalinowski:2018kdn,Guo-He:2020nok,Yang:2021hcu,Kalinowski:2020rmb,Melfo:2011ie,Abercrombie:2015wmb,Ilnicka:2015jba,Blinov:2015qva,Poulose:2016lvz,Hashemi:2016wup,Wan:2018eaz,Belyaev:2018ext,Belanger:2015kga,Dercks:2018wch,Lu:2019lok,Gil:2012ya,Swiezewska:2015paa,Blinov:2015vma,Huang:2017rzf,Belyaev:2016lok,Arhrib:2013ela,Eiteneuer:2017hoh,Ghosh:2021noq,Jueid:2020rek,Krawczyk:2013pea,Enberg:2013ara,Krawczyk:2013wya,Kanemura:2018esc,Hashemi:2015swh,Ahriche:2018ger,Ilnicka:2018def} .

After the electroweak symmetry breaking, the IDM possesses five physical scalars: One Higgs boson $h^0$ which is identified as
the 125 GeV SM Higgs, two new neutral physical scalars $H^0$, $A^0$ and 
two charged physical scalars $H^\pm$. 
Both $H^0$ and $A^0$ could be dark matter candidates. Obviously, the discovery of a 
charged scalar boson would be a clear sign of nonminimal Higgs sectors and precise knowledge
of its production properties would be useful to reconstruct the full Higgs potential. Since all new IDM scalars are odd under the $\zz_2$ symmetry, they must be produced in {pairs} at the colliders. 
Moreover, these new inert scalars only couple to the  electroweak gauge bosons and  the Higgs boson of the SM. Consequently, at the LHC they should be {pair-produced} via processes of the Drell-Yan type: $q\bar{q^{\prime}}\to W^{*}\to A^0H^{\pm}$, $q\bar{q^{\prime}}\to W^{*}\to H^0H^{\pm}$, and $q\bar{q}\to Z^{*}(\gamma)\to H^{+}H^{-}$, or from  $gg, q\bar{q}\to h^{0^*}\to H^{+}H^{-}$  \cite{Ghosh:2021noq,Kalinowski:2020rmb}, where   $\bar{q^{\prime}}$ represents a different quark flavor. Furthermore, the charged scalars in the IDM could also be produced  in the same-sign pair production process $pp\to H^{\pm}H^{\pm}jj$ \cite{Aiko:2019mww, Arhrib:2019ywg,Yang:2021hcu} or from vector boson fusion-like production $pp\to H^{\pm}H^{\mp}jj$ \cite{Kalinowski:2020rmb}. 
At lepton colliders, charged scalars can be produced via the pair production process $e^{+}e^{-}\to H^{+}H^{-}$ \cite{Kalinowski:2018ylg,Kalinowski:2018kdn,Aoki:2013lhm}. Future muon colliders \cite{Han:2021udl,Akeroyd:1999xf} may have better potential to cover a wide range of the mass of charged scalars. 

The discovery of a charged scalar would be clear evidence of beyond the SM physics.
In the case such a discovery happens in $e^+e^-$ collider, a subsequent precision measurement of its properties will be crucial to determine its nature. In order to obtain sufficient accuracy, one-loop corrections to the various charged scalar production  modes  and its couplings have to be considered. Recently, one loop radiative corrections to $e^+e^- \to Zh^0$ \cite{Abouabid:2020eik, Aiko:2021nkb, Ramsey-Musolf:2021ldh} and $e^+e^- \to H^0A^0$ \cite{Abouabid:2020eik}  in the IDM have been evaluated. \textcolor{black}{In addition, there are many other works dealing with radiative corrections in the IDM either at one-loop order \cite{Arhrib:2015hoa,Kanemura:2016sos,Abouabid:2020eik, Aiko:2021nkb, Ramsey-Musolf:2021ldh,Banerjee:2019luv,Banerjee:2021oxc,Banerjee:2021anv,Banerjee:2021xdp,Banerjee:2021hal,Banerjee:2016vrp,Kanemura:2019kjg} or beyond the one-loop level \cite{Braathen:2019zoh,Braathen:2019pxr, Senaha:2018xek}.}
 Full one-loop calculations to $e^{+}e^{-}\to H^{+}H^{-}$ in  nonsupersymmetric models such as  the 2HDM and some  supersymmetric models  like the real
 minimal supersymmetric standard model (MSSM) and complex MSSM  have been presented in Refs \cite{Arhrib:1994xp,Diaz:1995kc,Arhrib:1998gr,Beccaria:2005un,Beccaria:2002vd,Guasch:1999uh, Guasch:2001hk,Heinemeyer:2016wey}. For example, it has been found  in the complex MSSM \cite{Heinemeyer:2016wey}  that radiative corrections to $e^{+}e^{-}\to H^{+}H^{-}$   can go up to $20\%$ or higher, hence  a full one-loop contributions are important for future linear colliders such as the ILC or CLIC.

In this work,  we  present  for the first time a full one-loop analysis  to $ e^{+}e^{-}\to H^{+}H^{-}$ within the IDM. {Besides} the full weak corrections,  we  include the soft and hard QED radiation as well as the treatment of collinear divergences and the Coulomb singularity.
 In our scan and numerical analysis, we will take into account all the existing constraints on the IDM including the theoretical ones, the experimental constraints from the LHC, such as the Higgs data from the decay of the Higgs boson into two photons, the invisible Higgs decay, the electroweak precision  tests, as well as the constraints  from  dark  matter relic density in the Universe and the direct bounds from the monojet searches at the LHC. {It should be mentioned that we have included the recast results from LEP II data \cite{Lundstrom:2008ai} in our scan and analysis. Additional LHC recast results, such as dilepton and multilepton final states, as well as vector boson fusion final states, can be found in Refs. \cite{Belanger:2015kga,Dercks:2018wch,Belyaev:2022wrn,Robens:2019ynf}.  These recasts might bring new limits to the IDM parameter space. However, in this work, they are not taken into consideration. }
 
 We perform a scan over the whole parameter space of the IDM in three scenarios: Scenario I assumes that the inert scalars are degenerate in mass and has taken into account all the constraints except the invisible Higgs decay and dark matter constraints, scenario II is defined by imposing all the constraints without dark matter ones and with a nondegenerate spectrum for the new inert scalars and scenario III is the same as scenario II but with dark matter constraints. We find that the dark matter constraints can greatly reduce the number of points in the parameter space. Furthermore, it is found that QED corrections are rather small and radiative corrections are dominated by the weak interactions. The size of weak corrections depends on the coupling $\lambda_{h^0 H^+ H^-}$, which is further related to the mass of charged scalar. In scenario III we find that for $\sqrt{s}=250$ and $\sqrt{s}=500$ GeV, the weak corrections are around $-6\% \sim-5\%$ and $-10\% \sim -3\%$, respectively. While for $\sqrt{s}=1000$ GeV, the weak corrections can reach $-15\% \sim +25\%$. The new feature is that the weak corrections can be positive near the threshold when the charged scalar mass is larger than 470 GeV. We have proposed six benchmark points for future lepton collider searches.\\ 
The outline of this paper is  as follows: 
In Sec.\,\ref{sec:model},  we briefly describe the model, including  its mass spectra, key trilinear and quartic scalar couplings, and list various theoretical and experimental constraints that we will take into account in this work. In Sec.\,\ref{sec:LO}, we provide the leading-order (LO) formula for the cross sections of the $e^+ e^- \to  H^{+}H^{-} $ process, introduce the on-shell renormalization scheme for the IDM, and set up basic notations and conventions. Moving on, we study the one-loop contributions to the $e^+e^-\to H^{+}H^{-} $ process and examine the importance of soft and hard photon emission in order to guarantee the cancellation of the infrared (IR) divergences  at   the next-to-leading order (NLO)  calculation. Furthermore, we tackle the challenge posed by the Coulomb singularity using efficient resummation techniques. 
We present our numerical results in Sec.\,\ref{sec:results}. In Sec.\,\ref{sec:BPs}, we propose some benchmark points (BPs) and 
examine their radiative corrections for future $e^+e^-$ colliders. We end this work with discussions in Sec.\,\ref{sec:conclusions}.

\section{Review of the inert doublet model}
\label{sec:model}
\subsection{A brief introduction to IDM}
The IDM is one of the simplest extensions beyond the SM. This model has an extra doublet 
$H_2$ which is added to the scalar sector of the SM. This doublet does not generate any VEV and it does not have direct coupling to the fermions of the SM. An unbroken $\zz_2$ symmetry is imposed such that {fermions, gauge bosons, and the SM doublet are invariant while the 
additional scalar doublet is odd i.e.  $H_2 \to -H_2$ under this symmetry}. The parametrization of the two doublets is given by
\begin{eqnarray}
H_1 = \left(\begin{array}{c}
G^\pm \\
\frac{1}{\sqrt{2}}(v + h^0 + i G^0) \\
\end{array} \right)
\qquad , \qquad
 H_2 = \left( \begin{array}{c}
H^\pm\\
\frac{1}{\sqrt{2}}(H^0 + i A^0) \\
\end{array} \right) \,,
\end{eqnarray}
where $G^0$ and $G^\pm$ are the Goldstone bosons gauged out, after electroweak symmetry breaking,  by the longitudinal components of $W^\pm$ and $Z$, respectively. 
The $v$ denotes the VEV of the SM Higgs  doublet $H_1$. \\

Then the renormalizable scalar potential can be given as:
\begin{eqnarray}
V  &=&  \mu_1^2 |H_1|^2 + \mu_2^2 |H_2|^2  + \lambda_1 |H_1|^4
+ \lambda_2 |H_2|^4 +  \lambda_3 |H_1|^2 |H_2|^2 + \lambda_4
|H_1^\dagger H_2|^2 \nonumber \\
&+&\frac{\lambda_5}{2} \left\{ (H_1^\dagger H_2)^2 + {\rm H.c.} \right\}.
\label{potential}
\end{eqnarray}
Note that because of the exact $\zz_2$ symmetry, the above potential has no mixing terms like $\mu_{12}^2 (H_1^\dagger H_2 +
H.c)$.  In addition, since  the potential must be Hermitian, all $\lambda_i, i = 1, \cdots, 4$ are dimensionless and real whilst the phase of
$\lambda_5$ can be absorbed by a suitable redefinition
of the fields $H_1$ and $H_2$. After spontaneous symmetry breaking of the group $SU(2)_L \otimes U(1)_Y$
down to  $U(1)_{em}$, we have five physical scalars: $h^0$ which is the SM 125 GeV Higgs boson and four inert scalars: $H$, $A$, $H^\pm$.
Their masses are given by:
\begin{eqnarray}
&& m_{h^0}^2 = - 2 \mu_1^2 = 2 \lambda_1 v^2 \nonumber \\
&& m_{H^0}^2 = \mu_2^2 + \lambda_L v^2 \nonumber \\
&&  m_{A^0}^2 = \mu_2^2 + \lambda_S v^2 \nonumber \\
&&  m_{H^{\pm}}^2 = \mu_2^2 + \frac{1}{2} \lambda_3 v^2
\label{spect.IHDM}
\end{eqnarray}
where $\lambda_{L,S}$ are defined as
\begin{eqnarray}
\lambda_{L,S} &=& \frac{1}{2} (\lambda_3 + \lambda_4 \pm \lambda_5)
\end{eqnarray}
From above relations, one can easily find the expressions of $\lambda_i$ as a function of physical masses: 
\begin{eqnarray}
\lambda_1&=&\dfrac{m_{h^0}^2}{2v^2} \nonumber\\
\lambda_3&=&\dfrac{2(m_{H^\pm}^2-\mu_{2}^2)}{v^2} \nonumber\\
\lambda_4&=&\dfrac{(m_{H^0}^2+m_{A^0}^2 -2 m_{H^\pm} ^2)}{v^2} \nonumber\\
\lambda_5&=&\dfrac{(m_{H^0}^2-m_{A^0}^2)}{v^2} 
\label{eq:lams}
\end{eqnarray}
The IDM involves eight independent parameters: five $\lambda_{1,...,5}$, $\mu_{1}$,  $\mu_{2}$
and the VEV. One parameter can be eliminated by using the minimization condition, while the VEV is fixed by the $Z$ boson mass, fine-structure constant  and Fermi constant  $G_F$. Finally, we are left with six independent  parameters, which we choose as follows 
\begin{eqnarray}
 \{ \mu_2^2, \lambda_2, m_{h^0}, m_{H^\pm}, m_{H^0}, m_{A^0} \}
 \label{param.IHDM}
\end{eqnarray}

One alternative parametrization is to use $\lambda_L$ 
or $\lambda_S$ in place of $\mu_2^2$, as can be seen  from Eq.~(\ref{spect.IHDM}). 
The advantage of such parametrization is that it is  directly connected to the evaluation of the relic density.

For completeness, we list below the triple and quartic scalar couplings that are needed in our numerical analysis:
\begin{eqnarray}
&&h^0 H^\pm H^\mp = -v \lambda_3 \equiv v\lambda_{h^0H^\pm H^\mp}\equiv v\lambda_{h^0SS}\nonumber\\
&&H^\pm H^\mp H^\pm H^\mp   =-4\lambda_2\nonumber\\
&&G^\pm H^\mp H^0=\dfrac{v}{2}(\lambda_4 + \lambda_5) \nonumber\\
&&{G^\pm H^\mp A^0}=\pm \dfrac{v}{2}(-\lambda_4 + \lambda_5)
\label{scalar-coup}
\end{eqnarray}

It is clear that the triple coupling $h^0 H^\pm H^\mp$ is proportional to $\lambda_3$ which in turn is proportional to $(m_{H^\pm}^2-\mu_{2}^2)$ as given in Eq.~(\ref{eq:lams}).

\subsection{The IDM parameter space and constraints}\label{thexpcons}
In this work, we study in the same parameter space as in our previous work~\cite{Abouabid:2020eik}. The parameter space is obtained by scanning the whole space with certain theoretical and experimental constraints. The constraints used are summarized as follows:

\subsubsection{Theoretical constraints}
\label{sec:thecon}
The parameters of the IDM are subject to several theoretical  constraints  that we shall impose throughout our numerical analysis.
\begin{itemize}
	\item Perturbativity:
	
	We require that each of  the quartic couplings of the scalar potential in Eq.~(\ref{potential}) is perturbative:
	\begin{eqnarray}
	|\lambda_i| \le 8 \pi.
	\end{eqnarray}
	
	\item  Vacuum stability:
	
	In order to ensure vacuum stability, the potential $V$ should remain positive when the values of scalar fields 
	become extremely large \cite{Deshpande:1977rw}. From this condition, we have the following  set of constraints
	on the IDM parameters (for a review, see \cite{Branco:2011iw}):
	\begin{eqnarray}
	\lambda_{1,2} > 0 \quad \rm{and} \quad \lambda_3 + \lambda_4 -|\lambda_5| +
	2\sqrt{\lambda_1 \lambda_2} >0 \quad\rm{and} \quad\lambda_3+2\sqrt{\lambda_1
		\lambda_2} > 0.
	\end{eqnarray}
	
	\item  Charge-breaking minima:
	
	A neutral charge-conserving vacuum can be guaranteed by demanding that \cite{Ginzburg:2010wa}
	\begin{equation}
	\lambda_4-|\lambda_5|\leq 0.
	\label{chargebreaking}
	\end{equation}
	\item  Inert vacuum:
	
 We  impose the following conditions in order to insure that {the inert vacuum is the global one}~\cite{Ginzburg:2010wa}:
	\begin{eqnarray}
	m_{h^0}^2, m_{H^0}^2, m_{A^0}^2, m_{H^\pm}^2 >0 \qquad {\rm and} \qquad
	v^2>-
	\mu_2^2/\sqrt{\lambda_1\lambda_2}.
	\label{eq:inertvac}
	\end{eqnarray}
	\item Unitarity:
	
	As in the SM, the couplings of the IDM have to satisfy certain relations in order to obey unitarity. The tree-level perturbative unitarity is imposed on the various scattering 
	amplitudes of scalar bosons at high energy. From the technique developed in \cite{Lee:1977eg}, we find the following set of eigenvalues:
	\begin{eqnarray}
	&&e_{1,2}=\lambda_3 \pm \lambda_4 \quad , \quad
	e_{3,4}= \lambda_3 \pm \lambda_5\\
	&&e_{5,6}= \lambda_3+ 2 \lambda_4 \pm 3\lambda_5\quad , \quad
	e_{7,8}=-\lambda_1 - \lambda_2 \pm \sqrt{(\lambda_1 - \lambda_2)^2 + \lambda_4^2}
	\\
	&&
	e_{9,10}= -3\lambda_1 - 3\lambda_2 \pm \sqrt{9(\lambda_1 - \lambda_2)^2 + (2\lambda_3 +
		\lambda_4)^2}
	\\
	&&
	e_{11,12}=
	-\lambda_1 - \lambda_2 \pm \sqrt{(\lambda_1 - \lambda_2)^2 + \lambda_5^2}
	\end{eqnarray}
	
	We impose perturbative unitarity constraint on all $e_i$s: $e_i \le 8 \pi ~, \forall ~ i = 1,...,12$.
\end{itemize}
{It is important to note that we did not take into account higher-order corrections to the bounded from below, global minimum, and perturbative unitarity constraints. We have only included the constraints at lowest order. However, the parameter space will change if we use those constraints at NLO. For example, it has been shown in \cite{Ferreira:2015pfi} that the bounded from below conditions at one-loop level can change the stability of the electroweak vacuum, and the allowed parameter space for coexisting inert and inertlike minima at NLO is larger compared to the one observed at LO. Furthermore, for the unitarity constraints beyond the lowest order that have been analyzed in the 2HDM \cite{Cacchio:2016qyh}, it was found that perturbative NLO unitarity constraints are stronger than the LO ones. Therefore, the allowed ranges of the quartic $\lambda_i$  couplings, the inert scalar masses, and their mass differences will change if we use the NLO unitarity conditions.}

\subsubsection{Experimental constraints}
\label{sec:excon}
The parameter space of the scalar potential of the IDM should also satisfy 
experimental search constraints. We will consider the following experimental constraints (for further details about these constraints 
see our previous published work\cite{Abouabid:2020eik}):
\begin{itemize}
\item Higgs data at the LHC \cite{Aaboud:2018xdt,Sirunyan:2018ouh}. 
\item The bound on the invisible Higgs decay \cite{ATLAS:2020kdi}.
\item The direct collider searches from the LEP \cite{Arhrib:2012ia,Swiezewska:2012eh,Belanger:2015kga,Lundstrom:2008ai}
\item The indirect searches from electroweak precision tests \cite{Peskin:1991sw, Barbieri:2006dq, Tanabashi:2018oca}.
\item The data from dark matter searches \cite{Zyla:2020zbs, Abbott:1982af, Dine:1982ah, Preskill:1982cy,
 Belanger:2020gnr, Aprile:2018dbl, Amole:2019fdf, Abdelhameed:2019hmk, 
 Agnes:2018ves, Aaboud:2017phn, Sirunyan:2017hci, Belyaev:2018ext, Lu:2019lok}.
\end{itemize}

\section{Radiative corrections to $e^+ e^- \to H^{+}H^{-}$}
\label{sec:LO}
\subsection{$e^+ e^- \to H^{+}H^{-}$ at tree level}
\begin{figure}[!htb]\centering
\includegraphics[width=0.6\textwidth]{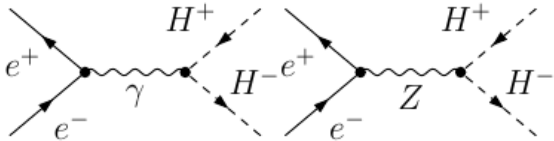}
	\caption{Tree level Feynman diagrams for $e^+ e^- \to H^+ H^-$ are shown.}
	\label{fig:eehphm}
\end{figure}

Due to the extremely small value of the electron mass and the corresponding Yukawa couplings, it is numerically justified to neglect the contributions proportional to the mass of electron, such as Feynman diagrams involving $e^{+}e^{-}h^0$, $e^{+}e^{-}G^{0}$, 
$e^{-}\overline{\nu_e}G^{+}$, and $e^{+}\nu_eG^{-}$ vertices.
For this reason, the Feynman diagrams contributing to $e^+e^-\rightarrow H^{+}H^{-}$ at the tree level are mediated by the $\gamma$ and 
$Z$ s-channel exchange as shown in Figure~\ref{fig:eehphm}.\\

Using the covariant derivative of the Higgs doublet, one can derive the scalar coupling to gauge bosons. We list below a part of the Lagrangian needed for our study 
\begin{eqnarray}
{\cal L}_{VH^+H^-,VV H^+H^-}&=&\biggl[-i e A^\mu + i e\frac{(c_W^{2}-s_W^{2})}{2c_W s_W} Z^\mu\biggr] H^+\stackrel{\leftrightarrow}{\partial}_\mu H^- \nonumber \\
&& +\biggl[e^2 A^\mu A_\mu - e^2\frac{(c_W^{2}-s_W^{2})}{c_W s_W} Z^\mu A_\mu \biggr] H^+ H^- \label{Lag}
\end{eqnarray}
where $e$ is the electric charge and $c_W\equiv\cos\,\theta_W$ ($s_W\equiv \sin\, \theta_W$) with  $\theta_W$ being the Weinberg mixing angle. 
It is clear from above equation that the interactions involving charged scalar 
depend only on the electric charge and  the Weinberg mixing angle $\theta_W$.

Let $p_{1,2}$ and $k_{1,2}$ be the momenta of incoming $e^\pm$ and  outgoing $H^\pm$, respectively,
and define the Mandelstam variables as
\begin{eqnarray}
& & s =  (p_1+p_2)^2 = (k_1+k_2)^2  \nonumber\\
& & t = (p_1-k_1)^2 = (p_2-k_2)^2   \nonumber\\
& & u = (p_1-k_2)^2 = (p_2-k_1)^2 
\end{eqnarray}
The matrix elements at tree level for the two contributions have the following form:
\begin{eqnarray}
{\cal M}_0^{\gamma} &=& -2\frac{e^2}{s}
\bar {v}(p_2)\not k_1 u(p_1)\label{2.28},\\
{\cal M}_0^{Z}  &=& 2 \frac{e^2g_H}{s-m_z^2} \bigm ( g_V \bar {v}(p_2)
\not k_1 u(p_1) -
g_A \bar {v}(p_2) \not k_1 \gamma ^5 u(p_1) \bigm )\label{2.29}
\end{eqnarray}
with
\begin{equation}
g_V=\frac{1-4 s_W^2}{4 c_W s_W}, \quad 
g_A=\frac{1}{4 c_W s_W}, \quad 
g_H=-\frac{c_W^2-s_W^2}{2 c_W s_W}.
\nonumber
\label{eq.phase}
\end{equation}
The total matrix element is  then given by
\begin{equation}
{\cal M}_{0}={\cal M}_0^{\gamma}+ {\cal M}_0^{Z}\label{30}
\end{equation} 
and the total cross section can be written as
\begin{equation}
\sigma^0=\frac{\pi\alpha^2\beta^3}{3 s}
\left[1+g_H^2 \frac{g_V^2+g_A^2}{(1-m_Z^2/s)^2} - 
\frac{2 g_H g_V}{1-m_Z^2/s}\right]\label{eq:LO},
\end{equation}
where $\beta=\sqrt{1-{4 m_{H^\pm}^2}/{s}}$ is the velocity of outgoing $H^\pm$ in the c.m. frame. Note that the first term comes from the photon exchange, the second term comes from the $Z$ boson exchange while the third one is the interference between the photon and $Z$ boson exchange. 

This cross section is only related to the mass of $H^{\pm}$ and it is independent of the other four parameters $\mu^{2}_2, \lambda_{2}, m_{H^0}, m_{A^0}$. It should be reminded that, near the threshold regions, it drops quickly due to the phase space suppression factor $\beta^3$. 

Lastly, as the collision energy is assumed to be 250 GeV or even higher, the intermediate $Z$ boson is always far away from its mass shell.  Hence, we have neglected the effects of decay width of $Z$ boson in this work.

\subsection{$e^+ e^- \to H^{+}H^{-}$ at one loop}
\label{sec:renormalization}
The 't Hooft-Feynman gauge has been used to evaluate both the weak corrections as well as the  QED ones.
The generic Feynman diagrams for $e^+ e^- \to H^{+}H^{-}$ are drawn in Fig.~\ref{figure-hphm}, which can be put into six categories: 
\begin{enumerate}
\item  One-loop corrections to the initial state vertices $e^+e^-V$ ($V=\gamma, Z$) which are purely SM, $G_{1}$  and $G_{2}$.\\
\item  One-loop corrections to the vertices $V H^+H^-$ ($V=\gamma, Z$), $G_3$ to $G_{11}$. \\
\item One-loop corrections to  photon and $Z$ boson  propagators as well as $\gamma$--$Z$, 
$G_{12}$ to $G_{18}$. Note that the mixing $Z$-$G^0$ and $\gamma$-$G^0$  
in the $s$-channel contribution was not included. Due to Lorentz invariance, such mixing would be proportional to $(p_1+p_2)^{\mu}$,
which after contracting with the initial state $e^+e^-$ will be proportional to $m_e$ and thereafter  will vanish. \\
\item  Box contributions, $G_{19}$  to $G_{21}$. \\
\item The various counterterms for initial and final states and also the $\gamma$ and $Z$
propagators, $\gamma-Z$  mixings are also depicted in $G_{22}$ to $G_{24}$.\\
\end{enumerate}

Moreover, we also add real photon emission $e^+ e^- \to H^{+}H^{-}\gamma$  which is depicted in diagrams  $G_{25}$  to $G_{29}$. 
Note that diagram $G_{27}$ does not contain IR divergence but it is an important piece for electromagnetic gauge invariance.

{
Let $M_V$ be the matrix element of all these diagrams,  the virtual corrections at NLO can be expressed as 
\begin{equation}
d\sigma_V \propto 2\mathrm{Re}\overline{\Sigma}\left(M_0M_V^\dagger\right). \nonumber
\end{equation}
Since we have omitted the decay width of the intermediate $Z$ boson in the leading order matrix element $M_0$, at fixed order we can only take into account the real part of $M_V$. This also means that we can neglect 
the imaginary parts of diagrams $G_{12}$  to $G_{18}$, which contribute to the decay width of the $Z$ boson at higher orders  when both vector bosons $V$ attached to the loop are $Z$ bosons.

\begin{figure}[!ht]
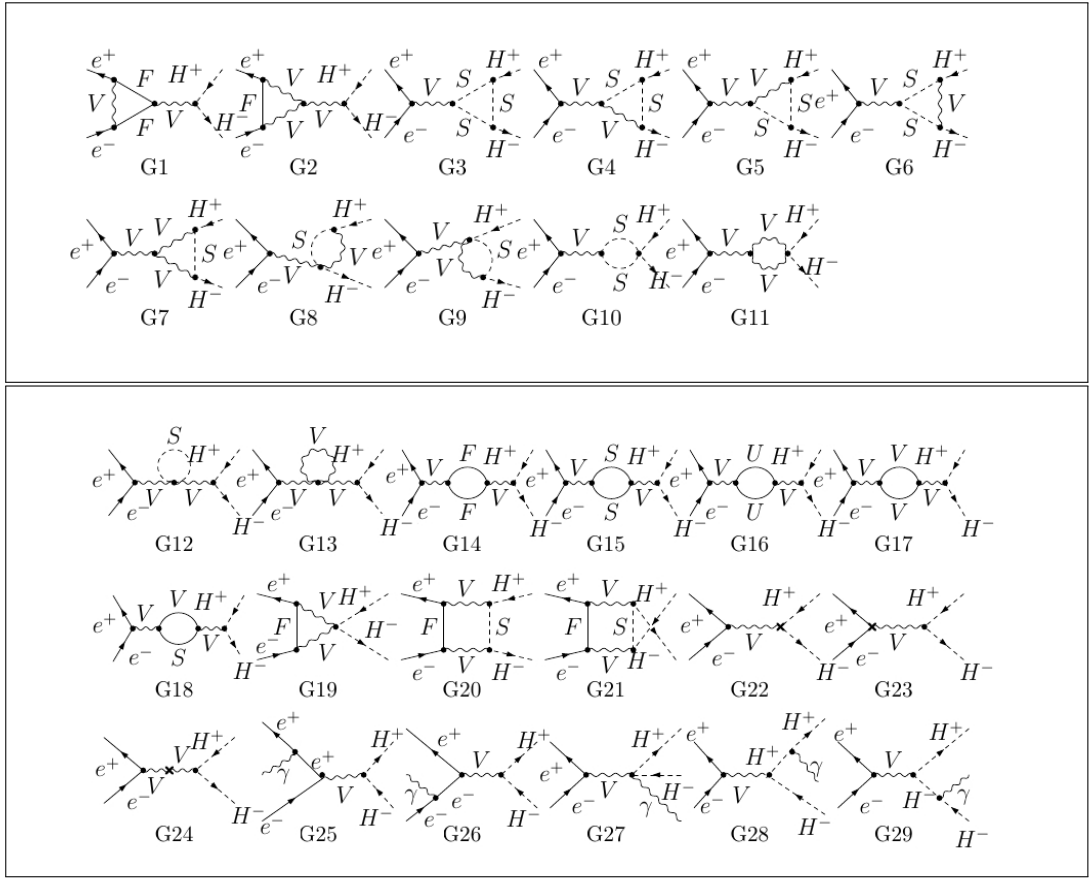

	\begin{center}
		\framebox[0.9\textwidth]{\includegraphics[width=0.8\textwidth]{vert1.pdf}}\\
		\framebox[0.9\textwidth]{\includegraphics[width=0.8\textwidth]{vert2.pdf}}
		\caption{Generic one-loop Feynman diagrams for $e^+e ^- \to H^+H^-$ are shown, 
			where $F$ stands for SM fermions; $V$ stands for generic vector bosons which could be $\gamma$, $Z$, or $W^\pm$; and 
			$S$ could be either a Goldstone $G^0$, $G^\pm$ or a Higgs boson $h^0,H^0,A^0$, or $H^\pm$.}
		\label{figure-hphm}
	\end{center}
\end{figure}

Calculation  of the one-loop corrections will lead to ultraviolet (UV)  as well as IR divergences. The UV singularities are treated in the on-shell renormalization scheme after being regularized using dimensional regularization
 while the IR divergences arising from the diagrams involving a photon are regularized with a small fictitious photon mass $\lambda$ and have to cancel with the ones from real photon emissions.

 The unbroken $\zz_{2}$ symmetry prevents the mixing between the SM doublet $H_1$ and the inert doublet $H_2$ which significantly facilitates the renormalization of the IDM.
The full renormalization of this model has been presented in~\cite{Banerjee:2019luv}. In our work, we will use the on-shell scheme developed first for the SM in \cite{Bohm:1986rj,Hollik:1988ii,Denner:1991kt}, completed by the on-shell renormalization  scheme for the inert scalar fields and their masses.  

Concerning the renormalization of the SM parameters and fields, 
we refer  to \cite{Denner:1991kt,Denner:2019vbn}.
For the renormalization of charge, an $\alpha(m_Z)$ scheme is used, following the procedure in Refs.~\cite{Denner:1991kt,Denner:2019vbn}.
The charge is first renormalized in the Thomson limit and then switched effectively to $m_Z$ by resumming large logarithms from light fermions, which gives
\begin{equation}
\alpha(m_Z)=\dfrac{\alpha(0)}{1-(\Delta\alpha(m_Z))_{f\neq\mathrm{top}}},
\label{eqn:renor}
\end{equation}
with
\begin{equation}
(\Delta\alpha(m_Z))_{f\neq\mathrm{top}}
=\Pi_{f\neq \mathrm{top}}(0)-\mathrm{Re}\Pi_{f\neq \mathrm{top}}(m_Z^2)
\end{equation}
and $\alpha(0)$ is the coupling constant renormalized in the Thomson limit. 
More details about this renormalization scheme can be found in Appendix~\ref{sec:charge}, where the uncertainties from different schemes are also discussed.
Following the notation in PDG~\cite{Tanabashi:2018oca}, Eq.~(\ref{eqn:renor}) is regarded as an ``on shell'' definition of the running coupling constant at the scale $m_Z$.

For the renormalization of the inert scalar bosons, we use a similar approach as in~\cite{Denner:1991kt,Denner:2019vbn}.
Because of  the exact $\zz_2$ symmetry, there is no mixing between $H^\pm$-$W^\mp$ and $H^\pm$-$G^\mp$, this makes the renormalization of the scalar fields easier.
The Higgs wave function renormalization constant and  mass counterterm are fixed by the two following on-shell conditions :
\begin{itemize}
	\item The on-shell condition for the physical mass $m_{H^{\pm}}$.
  \item The propagator of the charged scalar must have residue 1 at the pole mass.
\end{itemize}
These requirements will lead, respectively,  to
{
\begin{equation}
\begin{aligned}
\textrm{Re} \hat{\Sigma}_{{H^{\pm}}{H^{\pm}}}(m_{H^{\pm}}^2) &= 0   \\
\textrm{Re} \frac{\partial\hat{\Sigma}_{{H^{\pm}}{H^{\pm}}}(k^2)}{\partial k^{2}}\Bigg|_{k^2=m_{H^{\pm}}^2}& = 0 ,
\label{2HDM:reno}
\end{aligned}
\end{equation}
where $ \hat{\Sigma}_{{H^{\pm}}{H^{\pm}}} $ is the one-loop renormalized charged scalar self-energy, i.e.
\begin{equation}
 \hat{\Sigma}_{{H^{\pm}}{H^{\pm}}}(k^2)= \Sigma_{{H^{\pm}}{H^{\pm}}}(k^2)-\delta m_{H^\pm}^2 +(k^2-m_{H^\pm}^2)\delta Z_{H^\pm} .
 \label{eqn:rse}
\end{equation}
Here $\Sigma_{{H^{\pm}}{H^{\pm}}} $ is the one-loop unrenormalized charged scalar self energy.

 All the SM renormalization constants defined above can be found in  \cite{Denner:1991kt,Denner:2019vbn}. 
 Using the condition in Eq. (\ref{2HDM:reno}) and the relation in Eq.~(\ref{eqn:rse}), one can prove that
 \begin{eqnarray}
\delta Z_{H^\pm} = - \textrm{Re} \frac{\partial {\Sigma}_{{H^{\pm}}{H^{\pm}}}(k^2)}{\partial k^{2}}\Bigg|_{k^2=m_{H^{\pm}}^2}.
\end{eqnarray}
The explicit expression for  $\delta Z_{H^\pm} $ is presented in Appendix~\ref{sec:zh}.
}

For the counterterms, only two are new here comparing with the SM. They are listed as follows:
\begin{eqnarray}
&&\delta {\mathcal{L}}_{A H^{+} H^{-}}=-i e(\delta Z_{e}+\delta Z_{H^{\pm}}+\frac{\delta Z_{AA}}{2} + g_H \frac{\delta Z_{AZ}}{2})  A^\mu  H^+\stackrel{\leftrightarrow}{\partial}_\mu H^-  \\
&&\delta {\mathcal{L}}_{Z H^{+} H^{-}}=-i e g_H (\delta Z_{e}+\delta Z_{H^{\pm}}+\frac{\delta Z_{ZZ}}{2} -\frac{\delta{s_W}}{(c_W^{2}-s_W^{2})c_W^{2} s_W}
+\frac{\delta Z_{AZ}}{2g_H} ) 
Z^\mu  H^+\stackrel{\leftrightarrow}{\partial}_\mu H^-\nonumber
\label{eq:CT}
\end{eqnarray}
For the counterterms of the initial state vertices: $e^+e^-\gamma$ and  $e^+e^-Z$, counterterm of the $Z$ boson, the photon propagators, and their mixing , 
 they are exactly the same as in the SM and can be found in \cite{Denner:1991kt,Denner:2019vbn}. \\

{
With such a ``completely on-shell'' renormalization scheme, our results are totally independent of the renormalization scale $\mu_r$. As an alternative, they now depend on the new scale introduced during the renormalization of charge (set to $m_Z$ in this work).

As mentioned before, Feynman gauge is used in our calculation. Strictly speaking, to ensure our results are gauge invariant, calculation with another gauge is needed.
However this will lead to a totally different model file and/or modifications in the automatic package, which is beyond our abilities at present. Hence, we can only believe that this is automatically ensured by the theory, and only check those within our abilities (such as current conservation in real emissions). 

On the other hand, as already known in the renormalization of 2HDM, improper treatment of tadpole contributions may lead to gauge-dependent  results (see e.g. \cite{Denner:2016etu,Denner:2018opp,Krause:2016oke}). Fortunately, compared with the general 2HDM, the IDM is simple on this issue since there is no mixing between the two doublets. Hence, we can apply the same treatment as in the SM and avoid such an issue. 
}

Let us now discuss the treatment of the IR divergences, which includes two parts, one is from virtual corrections, the other is from real emission.
As mentioned before, IR divergences in this work are regularized by introducing a small fictitious photon mass, $\lambda$.
The IR divergences in virtual corrections are present in four sources: (i) wave function renormalization of charged particles such as electrons and charged scalars; (ii) vertex corrections to 
$e^+e^-\gamma$  and $e^+e^- Z$, $G_{1}$ in Fig.~\ref{figure-hphm} with $V=\gamma$,  where incoming electron and positron exchange a virtual photon with each other;
(iii) in the case of $e^+ e^- \to H^+H^-$ vertex corrections to $\gamma H^+ H^-$ and $ZH^+H^-$: $G_{6}$ in Fig.~\ref{figure-hphm} with $V=\gamma$ 
where the outgoing charged scalar pair exchanges a virtual photon ;
(iv)  in case of $e^+ e^- \to H^+H^-$ box corrections: $G_{20}$ and $G_{21}$ in Fig.~\ref{figure-hphm} with $(V,V)=(\gamma, \gamma)$ or $(V,V)=(\gamma, Z)$
where incoming fermions exchange one or two virtual photons with outgoing charged scalars.

IR divergence in real emission comes from phase space integration. 
Besides $\lambda$, two cutoffs $\Delta E$ and $\Delta\theta$ are introduced to deal with the IR singularities in real photon emission based on the two cutoff phase space slicing method~\cite{Harris:2001sx}. $\Delta E=\delta_s\sqrt{s}/{2}$ defines the soft photon energy cutoff for the bremsstrahlung process. It can be viewed as the photon energy cut that separates the soft and hard regions. 
The angle $\Delta \theta$ further separates the hard part into hard-collinear and hard-noncollinear regions.
The NLO corrections are then decomposed into the virtual, soft, hard-collinear, and hard noncollinear parts as follows:
\begin{equation}
d\sigma^{1}=d\sigma_V(\lambda)+d\sigma_S(\lambda,\Delta E)+d\sigma_{HC+CT}(\Delta E,\Delta\theta)+d\sigma_{H\overline{C}}(\Delta E,\Delta\theta) 
\label{fullcor}
\end{equation}
Here $S$, $HC$, and $H\overline{C}$ denote the contributions of soft, hard-collinear, and hard-noncollinear parts from real photon emission, respectively.
$V$ denotes the virtual corrections including loop and counterterm diagrams.
$CT$ in the third term of rhs denotes the extra contribution arising from the structure function of incoming electron and positron.
More details about the treatment of IR divergence are presented in Appendix \ref{sec:IR}, as well as the checks of our results for the independence of cutoffs.

The total cross section at NLO $\sigma^{\mathrm{NLO}}$, is the sum of LO cross section $\sigma^{0}$ and NLO corrections $\sigma^{1}$, namely
\begin{equation}
\sigma^{\mathrm{NLO}}=\sigma^{0} + \sigma^{1} \equiv \sigma^0(1+\Delta)\, ,
\label{sig1}
\end{equation}
where $\Delta$ is the relative correction.

As described in Sec.\, 3.1 of Ref.\cite{Beenakker:1991ca}, the NLO electroweak corrections can be safely grouped into two gauge-invariant parts:
\begin{enumerate}
\item The ``QED'' part, which includes all the diagrams that contain an extra photon attached to the LO diagrams. These diagrams can be easily found by investigating the generic Feynman diagrams in Fig.~\ref{figure-hphm} and we list them here:
\begin{itemize}
\item $G_1$ when the vector boson connected with the incoming electron pair is a photon;
\item $G_6$ when the vector boson connected with the outgoing charged scalar pair is a photon;
\item $G_8$ and $G_9$ when the vector boson connected with the outgoing charged scalar pair is a photon;
\item $G_{19}$, $G_{20}$, and $G_{21}$ when either vector boson is a photon; 
\item $G_{25}$ to $G_{29}$: all real emission diagrams. 
\end{itemize}
Meanwhile, the photonics contribution to the wave function renormalization of the electron and charged scalar is also grouped into this part.
\item The ``weak'' part, which contains all the remaining contributions.
\end{enumerate}
It should be noted that the ``QED'' part here is only a gauge-invariant subgroup of the whole QED corrections. 
There are remaining QED corrections in the ``weak'' part. 
Actually, it is very hard, if not impossible in this process, to separate the whole QED part because of $\gamma-Z$ mixing.

The relative correction is then separated into two parts correspondingly, i.e. we have
\begin{equation}
\Delta =\Delta_{\mathrm{weak}}+\Delta_{\mathrm{QED}}. 
\label{split}
\end{equation}

Similar to the pair production of $W$ bosons or top quarks, the Coulomb singularity appears in one-loop corrections, which is proportional to $\alpha\pi/\beta$ \cite{Fadin:1993kg,Bardin:1993mc}. 
It gives an important enhancement to the cross section near the threshold.
This effect is already well known and can be resummed to all orders.
{
After the resummation, the LO cross section becomes 
\begin{equation}
\sigma^0\rightarrow\sigma^0_{\mathrm{resum.}}=|\psi(0)|^2\sigma^0,
\label{eqn:resum0}
\end{equation}
where we have used the ``resum.''  as the abbreviation of resummed in the subscript to label the quantities after the resummation.
The factor $|\psi(0)|^2$, which was originally obtained by Sommerfeld~\cite{Sommerfeld:1931qaf} and Sakharov\cite{Sakharov:1948plh}, is given by 
\begin{equation}
|\psi(0)|^2=\dfrac{X}{1-e^{-X}}=1+\dfrac{X}{2}+\dfrac{X^2}{12}+\cdots
\label{eqn:expansion}
\end{equation}
where $X=\alpha\pi/\beta$.
}

{
Beyond LO, we assume that the resummed cross section can still be written into the product of two parts, similar to Eq.~(\ref{eqn:resum0}). The first part is  a factor that resums all the Coulomb singularities. Here we use $|\psi(0)|^2$ as this factor since we only resum the leading Coulomb singularities. The second component represents  a hard part which contains no more Coulomb singularities and can be calculated order by order perturbatively. Namely, we suppose
\begin{equation}
\sigma_{\mathrm{resum.}}=|\psi(0)|^2\sigma_{\mathrm{H}}.
\label{eqn:resum_all}
\end{equation}
Here the label ``H'' denotes the hard part. 
Expanding  the rhs of Eq.~(\ref{eqn:resum_all}) to LO in $\alpha$ and then matching it with LO cross section $\sigma^0$, it gives
\begin{equation}
\sigma^0_{\mathrm{H}}=\sigma^0.
\label{eqn:hard0}
\end{equation}
This  means that the LO resummed cross section in Eq.~(\ref{eqn:resum0}) is correctly reproduced. 
Now expanding the rhs of Eq.~(\ref{eqn:resum_all}) again to NLO in $\alpha$ and matching it with NLO cross section $\sigma^{\mathrm{NLO}}$, it gives
\begin{equation}
\sigma^0_{\mathrm{H}}+\dfrac{X}2\sigma^0_{\mathrm{H}}+\sigma^1_{\mathrm{H}}=\sigma^{\mathrm{NLO}}.
\label{eqn:hard10}
\end{equation}
Inserting Eqs.~(\ref{sig1}) and (\ref{eqn:hard0}) into Eq.~(\ref{eqn:hard10}) and solving $\sigma^1_{\mathrm{H}}$ from it, it gives
\begin{equation}
\sigma^1_{\mathrm{H}}=\sigma^0\left(\Delta-\dfrac{X}{2}\right) \equiv \sigma^0\Delta_{\mathrm{H}}.
\label{eqn:hard1}
\end{equation}
{Here, $\Delta_{\mathrm{H}}$ represents the ratio of $\sigma^1_{\mathrm{H}}$ to $\sigma^0_{\mathrm{H}}$. In our work, numerical analysis has confirmed that $\Delta_{\mathrm{H}}$ does not exhibit any further Coulomb singularities. This implies that Eq.~(\ref{eqn:resum_all}) works at least up to one-loop order. Then, the resummed cross section at NLO is derived as follows:}
\begin{equation}
\sigma_{\mathrm{resum.}}^{\mathrm{NLO}}=|\psi(0)|^2(\sigma_{\mathrm{H}}^0+\sigma_{\mathrm{H}}^1)=|\psi(0)|^2\sigma^0(1+\Delta_{\mathrm{H}}).
\label{eqn:resum1}
\end{equation}
Expanding $\sigma_{\mathrm{resum.}}^{\mathrm{NLO}}$ to even higher in $\alpha$, it gives
\begin{equation}
\sigma_{\mathrm{resum.}}^{\mathrm{NLO}}=\sigma_{\mathrm{H}}^0+\left(\sigma_{\mathrm{H}}^0\dfrac{X}{2}+\sigma_{\mathrm{H}}^1\right)+\left(\sigma_{\mathrm{H}}^0\dfrac{X^2}{12}+\sigma_{\mathrm{H}}^1\dfrac{X}{2}\right)+{\cal O}(\alpha^5).
\end{equation}
The first three terms on the rhs are the contributions of ${\cal O}(\alpha^2)$, ${\cal O}(\alpha^3)$,  and ${\cal O}(\alpha^4)$, respectively.
The third term stands for two different cases of Coulomb singularity at next-to-next-to-leading order. ${\sigma_{\mathrm{H}}^0X^2}/{12}$ corresponds to the exchange of two soft photons in the final states, while ${\sigma_{\mathrm{H}}^1X}/{2}$ corresponds to the exchange of one soft photon in the final states.
On the other hand, $\Delta_{\mathrm{H}}$ can also be separated into its QED and weak parts, similar to what is done for $\Delta$ in Eq.~(\ref{split}). This separation is expressed as
\begin{equation}
\Delta_{\mathrm{H}}=\Delta_{\mathrm{H, QED}}+\Delta_{\mathrm{H,weak}}.
\end{equation}
Due to the fact that Coulomb singularity is caused by soft photon exchange in the final states, it should belong to the QED part, namely we have
\begin{equation}
\Delta_{\mathrm{H,QED}}=\Delta_{\mathrm{QED}}-\dfrac{X}{2},\quad \Delta_{\mathrm{H,weak}}=\Delta_{\mathrm{weak}}.
\label{eqn:hardsplit}
\end{equation}
It has also been confirmed numerically that $\Delta_{\mathrm{H,QED}}$ remains finite when the velocity of outgoing charged pair goes to zero.\\
Up to this point,  we have effectively addressed the treatment of Coulomb singularities and obtained resummed cross sections.  However, in this work, we are more interested in relative corrections, not only the cross sections. Let $\Delta_{\mathrm{resum.}}$ be the ratio of  relative corrections after resummation ($\sigma_{\mathrm{resum.}}^{\mathrm{NLO}}/\sigma_{\mathrm{resum.}}^{0}-1$).
Due to the fact that $|\psi(0)|^2$ acts as a global factor in both LO and NLO cross sections, $\Delta_{\mathrm{resum.}}$ is always equal to the ratio of hard part, $\Delta_{\mathrm{H}}$. Correspondingly, we also have 
\begin{equation}
\Delta_{\mathrm{resum.,QED}}=\Delta_{\mathrm{H,QED}}=\Delta_{\mathrm{QED}}-\dfrac{X}{2}, \quad 
\Delta_{\mathrm{resum.,weak}}=\Delta_{\mathrm{H,weak}}=\Delta_{\mathrm{weak}} .
\label{eqn:resumsplit}
\end{equation}

\begin{table}[!htb]
\centering
\begin{tabular}{|c|rrrrrrr|}
\hline\hline
$m_{H^\pm}$ (GeV) &100&150&200&225&245&249&249.9 \\
\hline
$\beta$&0.9165&0.8000&0.6000&0.4359&0.1990&0.0894&0.0283\\
\hline
$|\psi(0)|^2$ &1.0134&1.0153&1.0204&1.0282&1.0625&1.1425&1.4918\\ 
\hline\hline
\multicolumn{8}{|c|}{Before Resummation} \\ 
\hline
 $\sigma^0$ (fb)                            &95.320      &63.392     &26.744   &10.254  &0.976  &0.0883 &0.00280\\
$\sigma^{\mathrm{NLO}}$ (fb)       &90.344     &60.288    &25.783     &10.085  &1.004  &0.0970 &0.00389 \\
 $\Delta_{\mathrm{QED}}$(\%)      &1.628       &1.711       &2.101      &2.731    &5.738  &12.911 &42.143\\
 $\Delta_{\mathrm{weak}}$(\%)     &-6.859      &-6.609      &-5.695    &-4.379   &-2.869 &-3.058 &-3.175\\
\hline\hline
\multicolumn{8}{|c|}{After Resummation} \\ \hline
 $\sigma^0_{\mathrm{resum.}}$ (fb)                            &96.593      &64.362     &27.291   &10.543  &1.037  &0.1009 &0.00418\\
 $\sigma^{\mathrm{NLO}}_{\mathrm{resum.}}$ (fb)       &90.256     &60.230   &25.756     &10.075  &1.003  &0.0971 &0.00401 \\
 $\Delta_{\mathrm{resum.,QED}}$(\%)      &0.299       &0.189      &0.071      &-0.064   &-0.384  &-0.723 &-0.933 \\
 $\Delta_{\mathrm{resum.,weak}}$(\%)     &-6.859      &-6.609      &-5.695    &-4.379   &-2.869 &-3.058 &-3.175\\
\hline\hline
\end{tabular}
\caption{Cross section and relative corrections before (second block) and after (third block) the resummation of the Coulomb singularity with $\sqrt{s}=500$ GeV {are demonstrated}.  The remaining IDM parameters are chosen as $m_{H^0}=m_{A^0}=m_{H^\pm}$, $\lambda_2=2$, and $\mu_2^2=0$. 
{Quantities before resummation, $\sigma^0$, $\sigma^{\mathrm{NLO}}$, $\Delta_{\mathrm{QED}}$, and $\Delta_{\mathrm{weak}}$, are obtained with Eqs.~(\ref{eq:LO}), (\ref{sig1}), and (\ref{split}). 
Quantities after resummation, $\sigma^0_{\mathrm{resum.}}$, $\sigma^{\mathrm{NLO}}_{\mathrm{resum.}}$, $\Delta_{\mathrm{resum.,QED}}$, and $\Delta_{\mathrm{resum.,weak}}$, are obtained with Eqs.~(\ref{eqn:resum0}), (\ref{eqn:hard1}), (\ref{eqn:resum1}), and (\ref{eqn:resumsplit}).} 
}
\label{Tab:resum}
\end{table}
In order to show the effect of this resummation technique, we present in a certain case the cross section and relative corrections before and after the resummation in Table~\ref{Tab:resum} .
There are a few comments on the results in the table:
\begin{itemize}
\item The relative weak corrections remain constant both before and after resummation, meaning that $\Delta_{\mathrm{resum.,weak}}$ maintains its equality to $\Delta_{\mathrm{weak}}$ throughout.
\item The resummation mostly affects the results near the threshold, i.e. the region where $\beta$ is small and close to 0. When $\beta$ is varying from $0.9165$ to $0.0283$, it is observed that the QED corrections can change from $1.628\%$ to $42.143\%$.
\item Before the resummation, the LO cross section can be changed drastically by the QED corrections, especially near the threshold region. After the resummation, the QED corrections are small and well controlled within a size around $1\%$. 
The LO cross section is enhanced directly by the factor $|\psi(0)|^2$ during the resummation, which varies from 
$1.0134$ to $1.4918$ as $\beta$ decreases.
While the NLO cross section is much more stable in this process. From the data before and after the resummation we can see the ratio is always  around $1$ (from $0.999$ to $1.039$ as $\beta$ decreases).

\item Throughout the entire region, the QED corrections decrease by at least one order of magnitude after resummation. This observation strongly suggests that the dominant factor contributing to the QED corrections is the Coulomb singularity term.
\end{itemize}
All of this demonstrates that the resummation is necessary and effective. In the following discussion, we will use $\sigma$ or $\Delta$ to refer to the resummed values, and for convenience, we will omit the subscript ``resum.''
}
\section{Numerical results}
\label{sec:results}
In the present work, the computation of all the one-loop matrix elements and
counter-terms {is performed} with the help of \texttt{FeynArts} and \texttt{FormCalc}~\cite{Hahn:2000kx,Hahn:1998yk,Hahn:2006qw}
packages. {The scalar integrals are numerically evaluated using
\texttt{LoopTools}~\cite{Hahn:1999mt,Hahn:2010zi}}. The other parts are obtained with the help of \texttt{FDC}~\cite{Wang:2004du} and \texttt{BASES}~\cite{Kawabata:1995th}. 
The cancellation of UV divergences are obtained both analytically and numerically.
It should be noted that the model file of IDM, including all needed renormalization counter terms and renormalization constants, is obtained by manually modifying the SM model file. We have checked this model file with the output from \texttt{FeynRules} and confirmed their agreement. Recently,  an automatic tool \texttt{NLOCT}~\cite{Degrande:2014vpa}, which can automatically generate counter terms and calculate renormalization constants, became available. It provides a model for 2HDM that  can be used in one-loop QCD and EW calculation. However, compared with the general 2HDM, IDM is simple because there is no mixing between the two doublets. Hence, we use our own model file mentioned above.

In this section, we present our numerical results for  our  process $e^+e^- \to H^+H^-$. It has been pointed out that the triple Higgs couplings can cause a large (can go up to $30\%$) radiative correction to the process $e^+e^- \to H^+ H^-$ \cite{Arhrib:1998gr,Guasch:1999uh} in the 2HDM. Below we examine the radiative corrections to our process  in the IDM and we expect that the electroweak radiative corrections to $e^+e^- \to H^+ H^-$ would have some similarities with $e^+ e^- \to H^0 A^0$ \cite{Abouabid:2020eik}. For the QED corrections, we will study  both soft photon emission as well as the hard one.

\subsection{Numerical input}
We adopt the following numerical values of the physical parameters from PDG \cite{Tanabashi:2018oca}:
\begin{enumerate}
	\item The fine structure constant: $\alpha(0)=1/{137.036}$, $\alpha(m_Z)={1}/{128.943}$ with $\Delta \alpha_{\textrm{hadron}}^{(5)}(m_Z) = 0.02764$.
	\item The gauge boson masses: $m_{W}=80.379$ GeV and {$m_{Z}=91.188$ GeV}.
	\item The fermion masses:
	\[ \begin{array}{lll}%
		m_{e}=0.511\, \text{MeV}  & m_u=0.134\, \text{GeV}  &  m_{t}=173.2\, \text{GeV},\\
		m_{\mu}=0.106\, \text{GeV}  &  m_d=0.134\, \text{GeV} &  m_b=4.660\, \text{GeV},\\
		m_{\tau}=1.777\, \text{GeV} & m_s=0.095 \, \text{GeV} &  m_c=1.275\, \text{GeV}.
	\end{array}\]%
	 The masses of $u$ and $d$ are taken as effective parameters to reproduce $\Delta \alpha_{\textrm{hadron}}^{(5)}(m_Z)$ with $\alpha(m_Z)$. They are obtained as $ m_u=m_d=0.134$ GeV. 
\end{enumerate}

In the IDM, the $CP$ even Higgs boson $h^0$ is the only  SM-like Higgs boson observed  by the LHC collaborations, and we use $m_{h^0}=125.18$ GeV. For the other IDM parameters, we perform a scan over the whole parameter space, which includes the physical masses $m_{A^0}$, $m_{H^0}$ and $m_{H\pm}$, $\mu_{2}^2$ and $\lambda_2$ parameters. We take into account all the experimental constraints as well as the theoretical requirements given in the Sec. \ref{thexpcons}.

It is found that our numerical results are almost independent of the $\lambda_{2}$ parameter.  Consequently, we will fix $\lambda_{2}= 2$ in the next part.

In the following,  we will use the $\alpha(m_Z)$ scheme described before to present our results. 

\subsection{Weak corrections in three scenarios}
In our numerical analysis, we will consider three scenarios, which are tabulated in Table \ref{tab:scenarios}. They are categorized in terms of degenerate/nondegenerate between inert scalar boson masses, with/without  invisible Higgs decay, and with/without DM constraints. It is thought that scenario I is the simplest one with only three free parameters  and is the easiest one to demonstrate the one-loop corrections, while scenario III might be more realistic after taking into account collider experimental bounds and  dark matter constraints and six benchmark points are chosen from this scenario for the LHC and future collider searches.

\begin{table}[!htb]
	\centering
	\begin{tabular}{|c|c|c|c|c|c|}
		\hline
		& Scenario I & Scenario II & Scenario III  \\
		\hline
		Theoretical constraints & $\checkmark$ & $\checkmark$& $\checkmark$\\
		\hline 
		Degenerate spectrum &$\checkmark$&&\\
		\hline
		\hline
		Higgs data &$\checkmark$&$\checkmark$&$\checkmark$\\
		\hline
		Higgs invisible decay open &&$\checkmark$&$\checkmark$\\
		\hline
		Direct searches from LEP &$\checkmark$&$\checkmark$&$\checkmark$ \\
		\hline
		Electroweak precision tests & $\checkmark$&$\checkmark$&$\checkmark$\\
		\hline
		Dark matter constraints & & &$\checkmark$\\
		\hline
	\end{tabular}
	\caption{Scenarios and their conditions are tabulated. Scenario I where all the inert scalars are degenerate is described by three parameters which are $(m_S=m_{H^0}=m_{A^0}=m_{H\pm},\mu_2^2,\lambda_{2})$, scenarios II and III with a nondegenerate spectrum for the inert scalars are described by five parameters, which are  $(m_{H^\pm}, m_{H^0}, m_{A^0},\mu_2^2, \lambda_2)$.}
	\label{tab:scenarios}
\end{table}

Our numerical analysis begins with scenario I, which has only three free parameters to consider. 
When $\lambda_2$ is fixed, there are only two free parameters ($\mu^2_2$ and $m_S=m_{H^0}=m_{A^0}=m_{H\pm}$) to vary. Three typical collision energies of future $e^+e^-$ colliders, namely: $\sqrt{s}=250$ GeV, $\sqrt{s}=500$ GeV and $\sqrt{s}=1000$ GeV are chosen to expose the effects of new physics.

It should be mentioned that when the charged scalar mass is fixed, {the $h^0H^+H^-$ coupling} is simply determined by the parameter $\mu_2^2$, as given in Eq. (\ref{eq:lams}). Therefore, in order to examine how the theoretical parameters like $\mu_2^2$ and $m_{H^\pm}$ can affect the cross section  and how theoretical constraints and experimental bounds can affect allowed parameter space, we introduce five cases with different values of $\mu_2^2$ given in Table~\ref{tab:idms}. These values of $\mu_2^2$ corresponding to five cases are chosen in the allowed parameter space which have passed all theoretical and experimental bounds and constraints, and we have labeled these results by the labels IDM1-5 in Fig. \ref{eehphm-degen}. We have deliberately introduced two with positive values and two with negative values of $\mu^2_2$.

	\begin{table}[!htb]
	\centering
	\begin{tabular}{|c|c|c|c|c|c|}
		\hline
		& IDM1  & IDM2 & IDM3 & IDM4 & IDM5  \\
		\hline
		$\mu_2^2$(GeV$^2$)  & 40000 & 6000  & 0  & -10000 & -30000 \\
		\hline
	\end{tabular}
	\caption{In scenario I, five cases with typical values of $\mu_2^2$  labeled as IDM1-5 are given.}
	\label{tab:idms}
\end{table}

In Fig.~\ref{eehphm-degen}, total cross section and relative corrections to $e^+e^-\to H^+ H^-$ as a function of the inert scalar masses $m_S$ are shown in scenario I. It is observed that the allowed ranges of $m_S$ are different when $\mu_2^2$ takes different values. 
{More explicitly, it is found that in the range of 100$-$500 GeV, the lower bounds of $m_S$ in the cases of IDM1, IDM4 and IDM5 are constrained by the signal strength of $h^0\rightarrow \gamma\gamma$ at the LHC, whereas the upper bound of $m_S$ in the case of IDM5 is constrained by the unitarity condition.}
For example, in the IDM5 case, $m_S$ shall not be smaller than 340 GeV, which explains why there is no IDM5 in the plot with $\sqrt{s}=500$ GeV and there are no IDM4 and IDM5 in the plot with $\sqrt{s}=250$ GeV. 

Obviously, since the process $e^+e^- \to H^+ H^-$ is $s$-channel dominant, when the mass of charged scalar is fixed, the cross section becomes smaller and smaller when the  c.m. energy increases from 250 GeV to 1 TeV.
For a fixed c.m. energy, the cross section drops quickly when the mass of charged scalar increases to the kinematic end points. It is noteworthy that a pair of light charged scalar (say 100 GeV or so)  can be copiously produced and the total cross sections could be of the order of 100 fb which would lead to more than $2 \times 10^5$ events at LC machines with $\sqrt{s}=250$ GeV which can deliver about 2 ab$^{-1}$ luminosity.\\
	\begin{figure}[H]
		\includegraphics[width=0.3\textwidth]{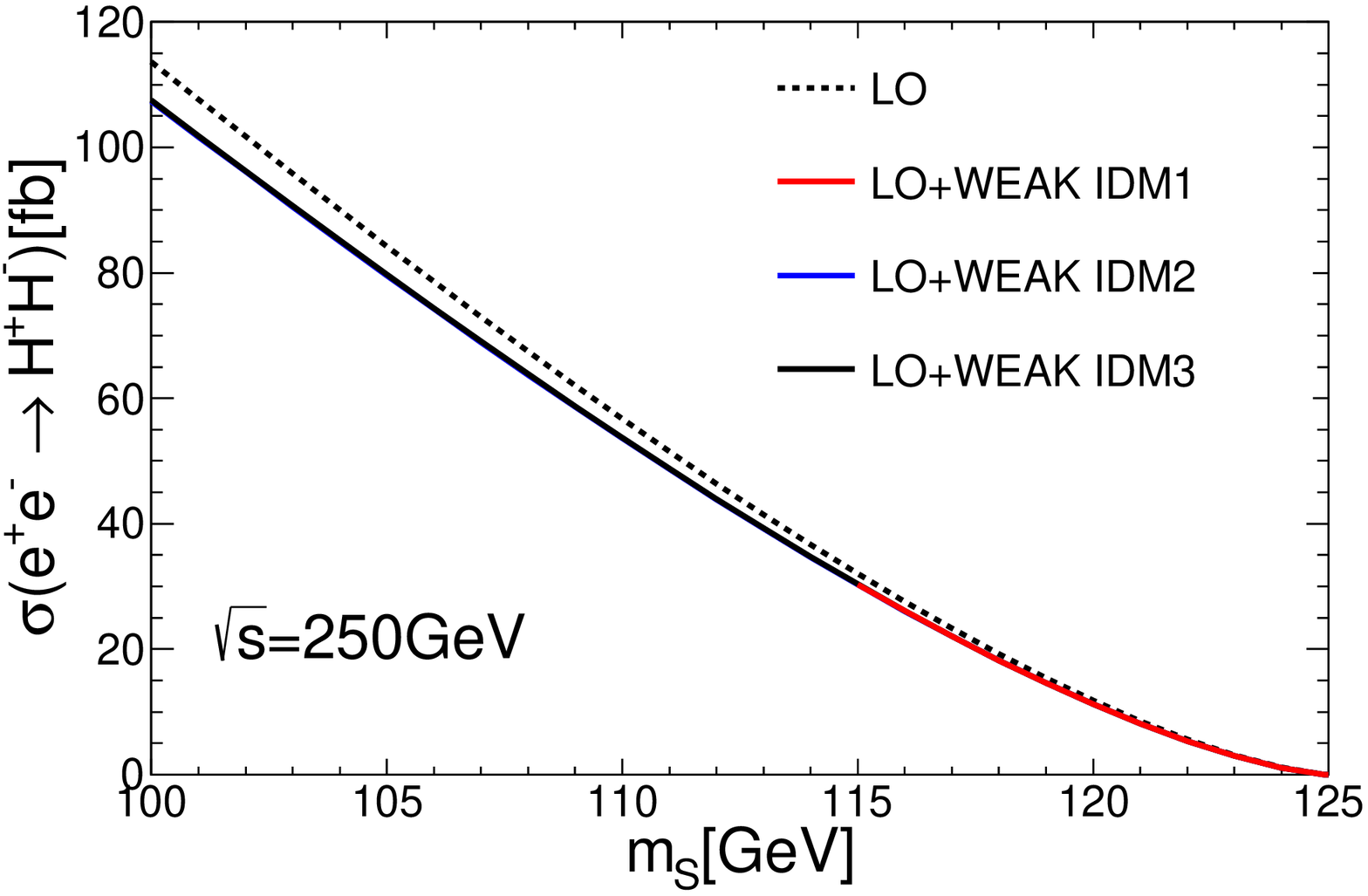}
		\includegraphics[width=0.3\textwidth]{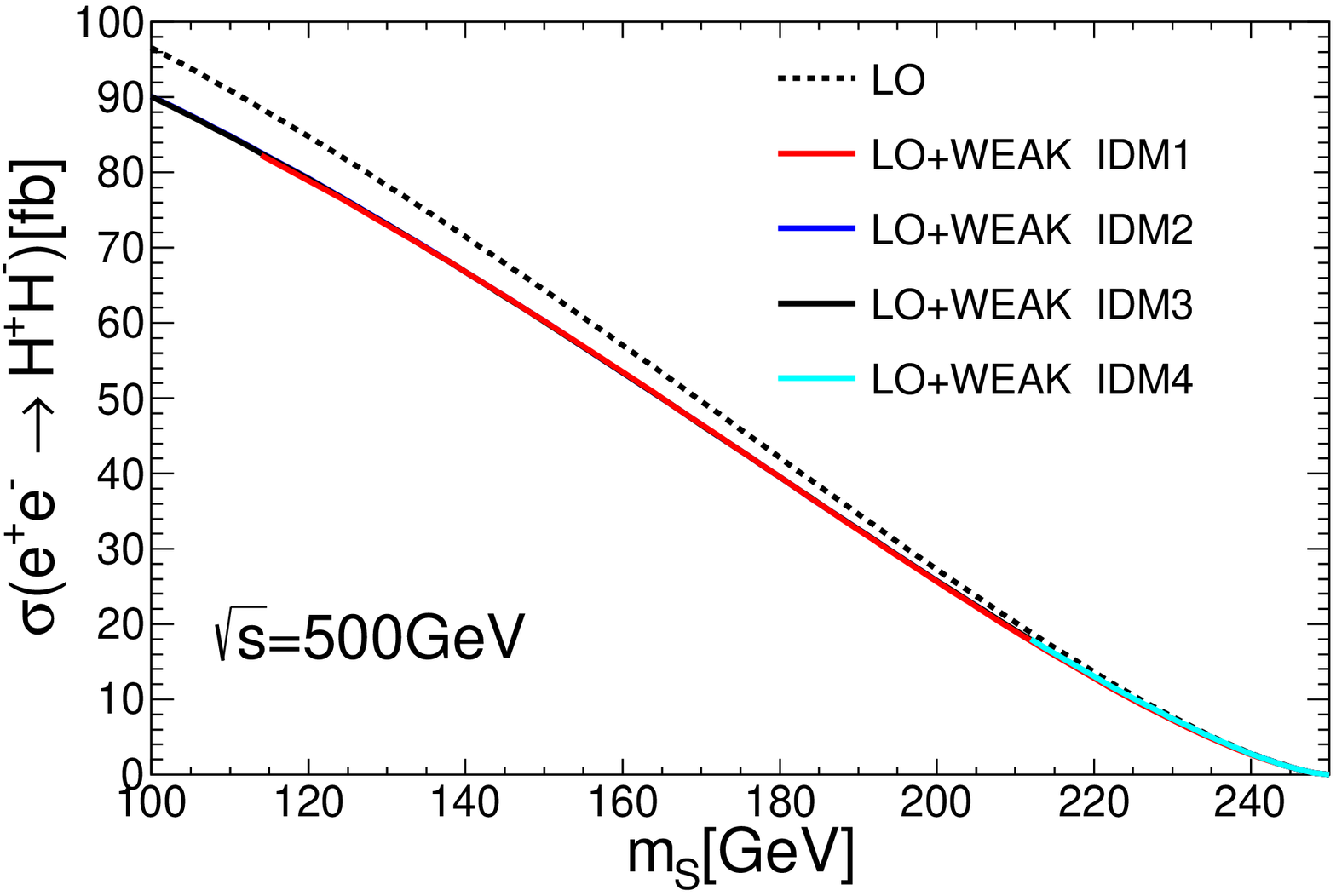}
		\includegraphics[width=0.3\textwidth]{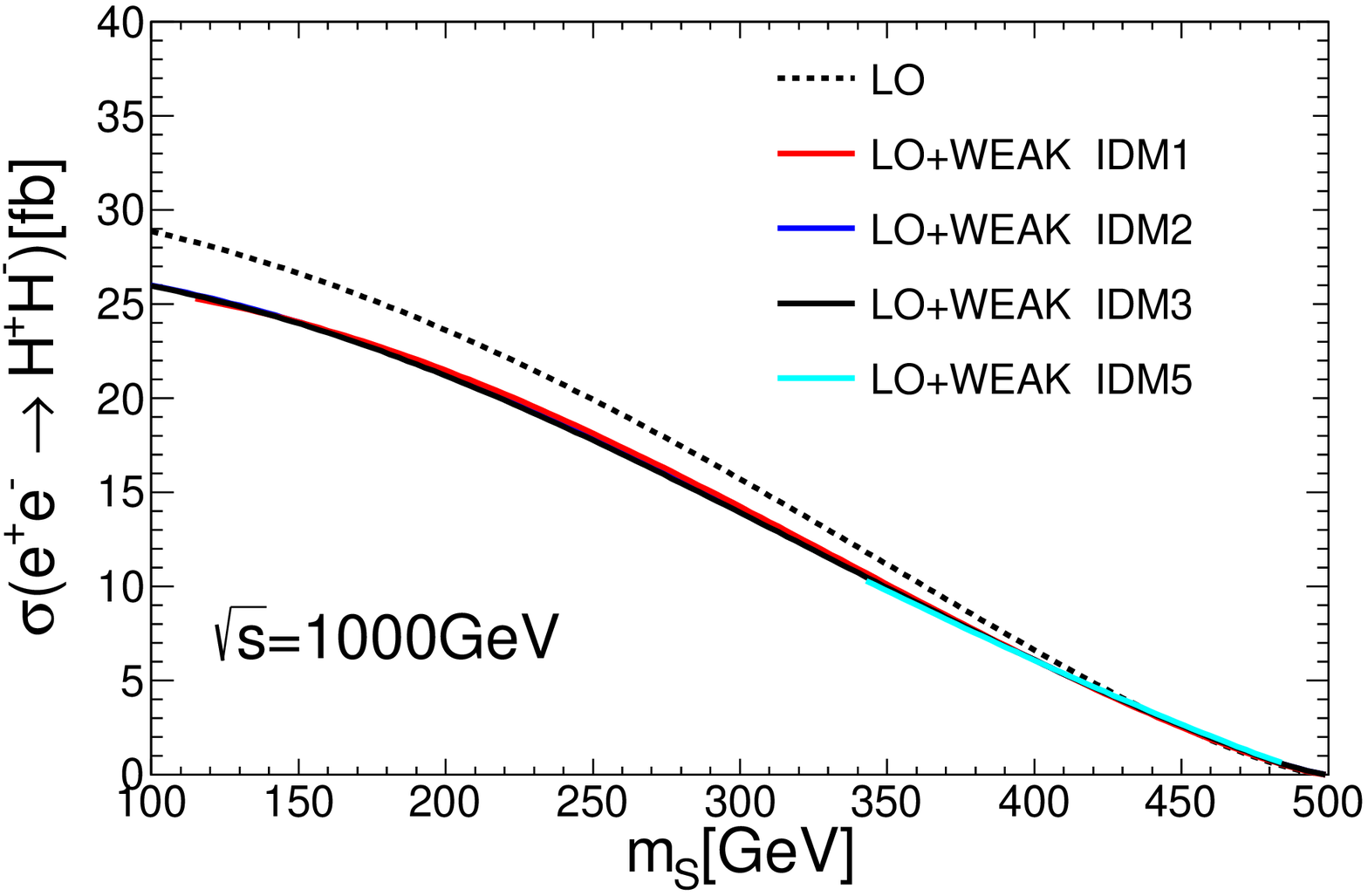}\\
		\includegraphics[width=0.3\textwidth]{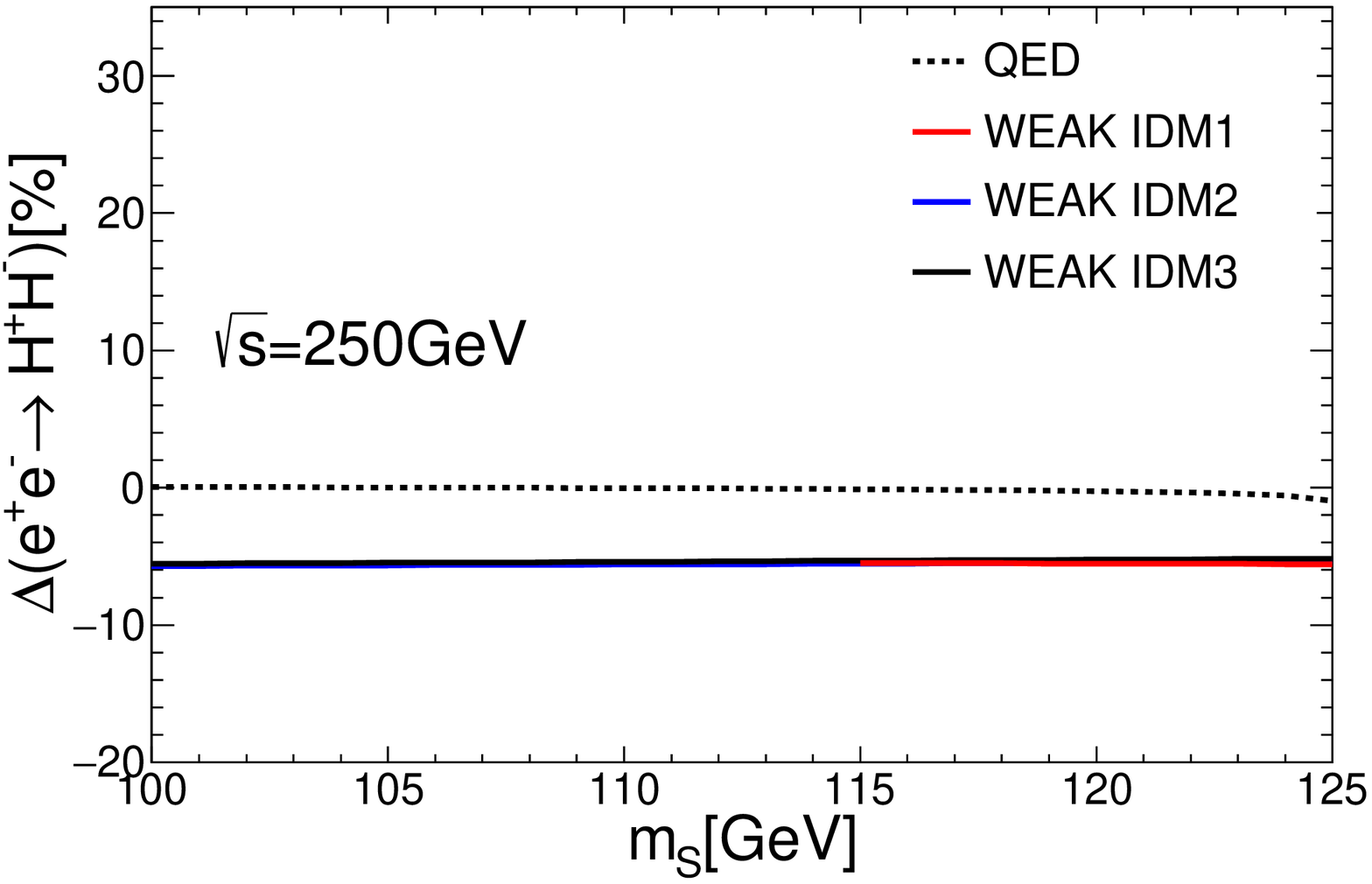}
		\includegraphics[width=0.3\textwidth]{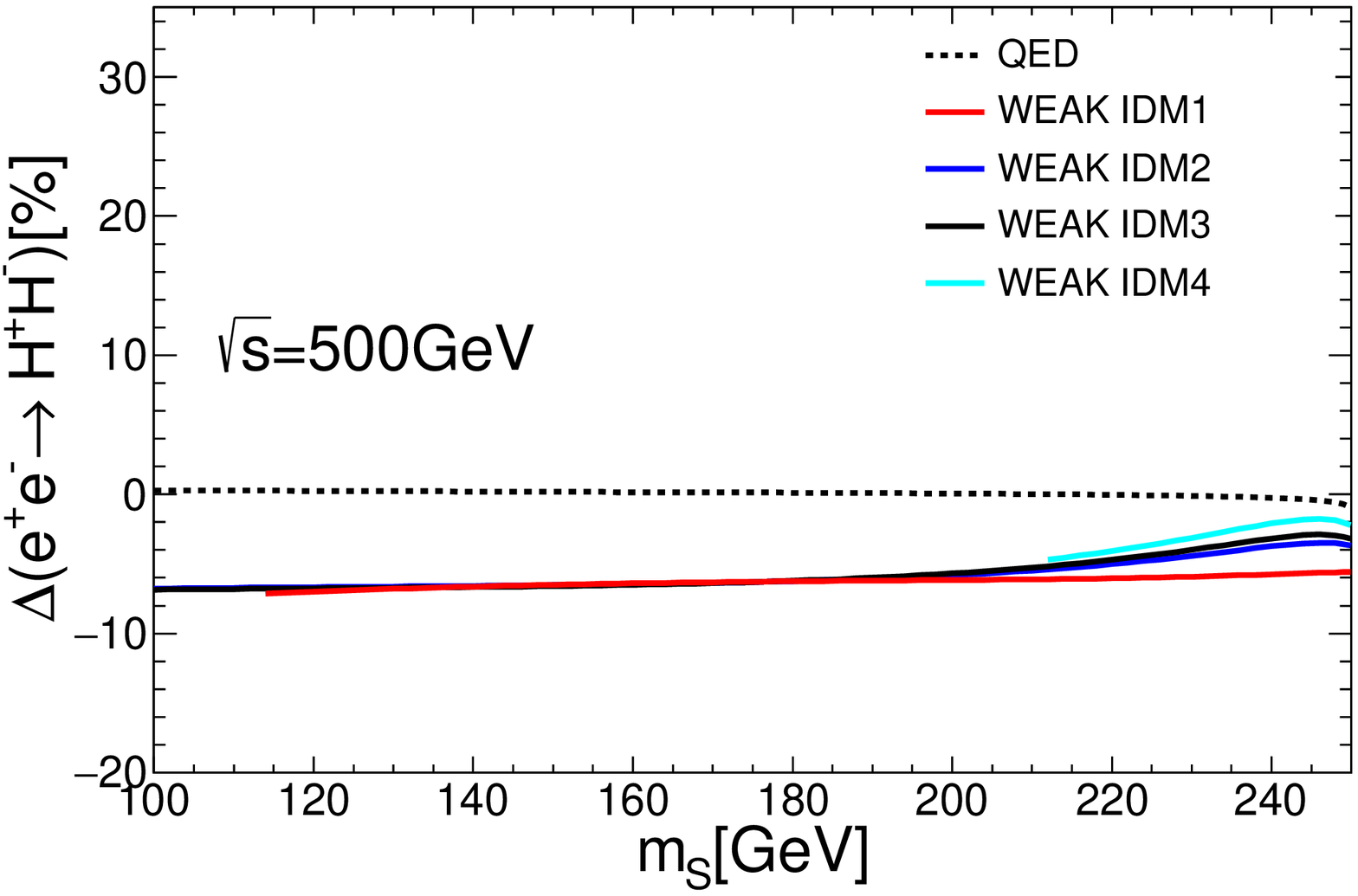}
		\includegraphics[width=0.3\textwidth]{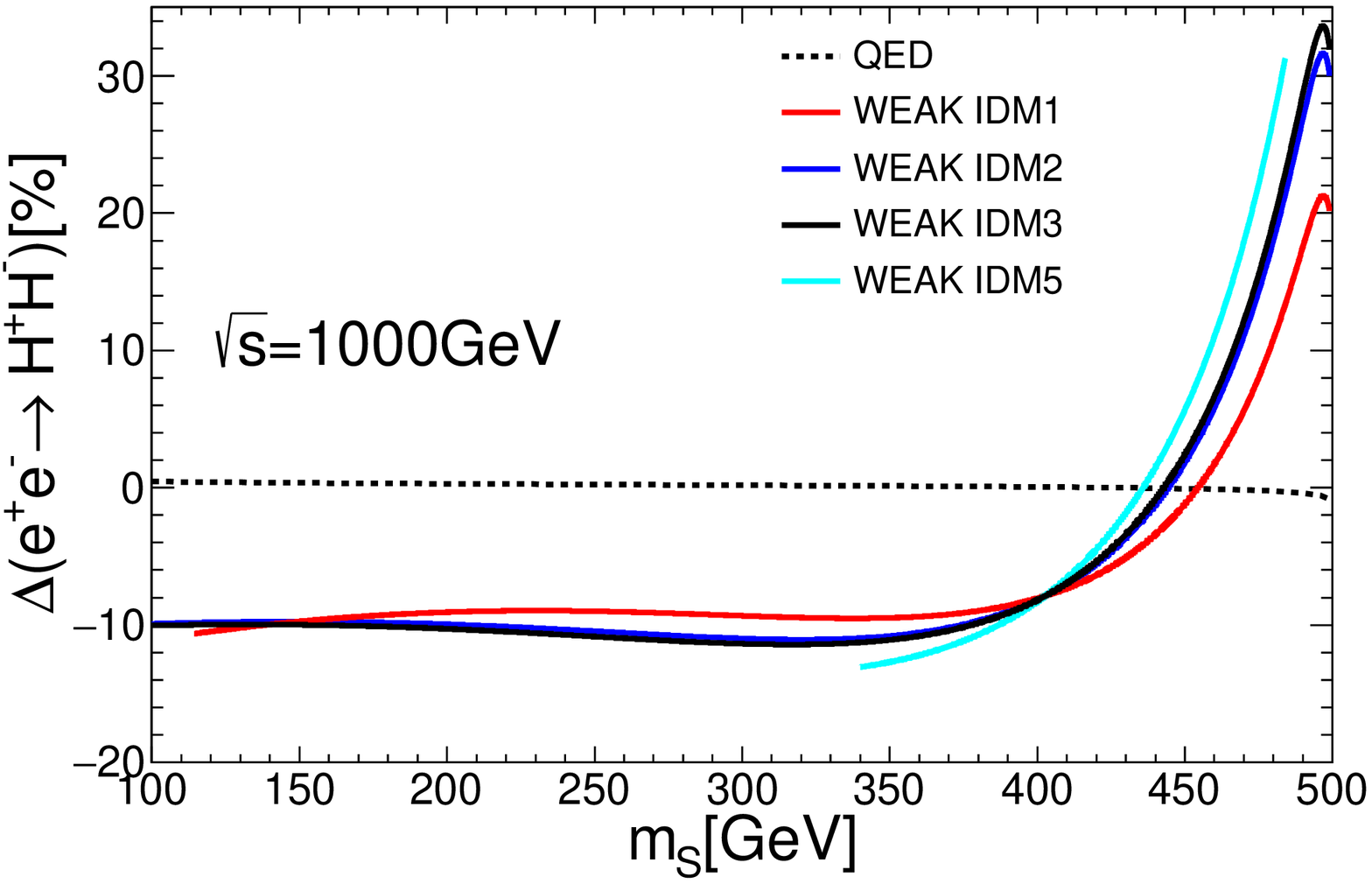}
		\caption{The total cross section and the ratio of EW corrections to $e^+ e^-\longrightarrow H^{+}H^{-}$ as a function of the inert scalar masses $m_S$ with three collision energies  $\sqrt{s}=250$ GeV, 500 GeV, and 1000 GeV  are shown in scenario I in the upper and lower panels, respectively. The corresponding values of $\mu_2^2$ are given in Table~\ref{tab:idms}.}
		\label{eehphm-degen}
	\end{figure}
	
It can be read from the lower panel of Fig.~\ref{eehphm-degen}  that the ratio of weak corrections in the IDM can reach $-6\%$ at $\sqrt{s}=250$ GeV, which is almost independent of the mass $m_S$. In contrast, at the c.m. energy $\sqrt{s}=1$ TeV,  the ratio starts from $-10\%$ and can go up to $30\%$ or higher. Such a behavior can be attributed to the large corrections of the $h^0H^+H^-$ coupling. 
	For the degenerate scenario with $\mu_2^2$ fixed, when the parameter $m_S=m_{H^\pm}$ increases from 100 GeV to 500 GeV or so, {the $h^0H^+H^-$ coupling} proportional to $\lambda_3$ also increases almost monotonically, especially near the end point region (say 350 GeV $<m_S<500$ GeV). Such a $h^0H^+H^-$ coupling can contribute to the EW corrections both linearly via the virtual corrections  to $(\gamma,Z )H^+H^-$ vertices and quadratically via the wave function renormalization of the charged scalar. When $m_S$ increases, the total cross sections decrease due to the suppression of phase space, while the ratio of EW corrections becomes larger and larger, as shown in the last plot of the lower panel.

\begin{figure}[h!]\centering
	\includegraphics[width=0.31\textwidth]{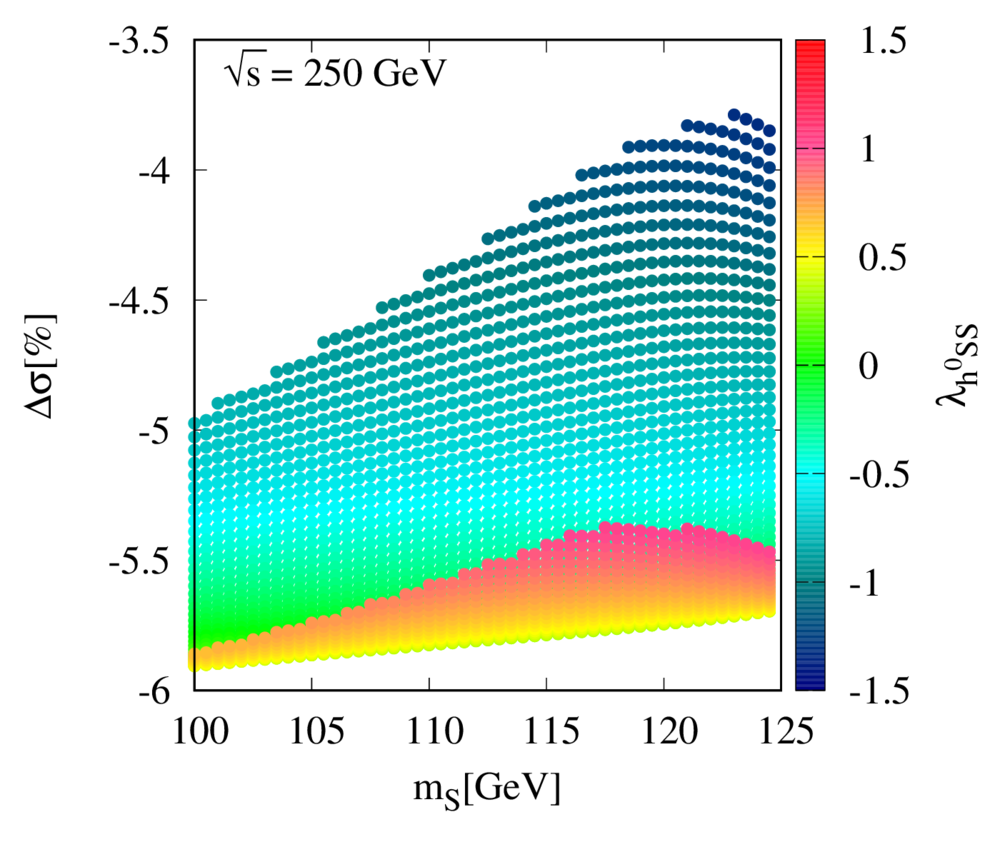}\includegraphics[width=0.31\textwidth]{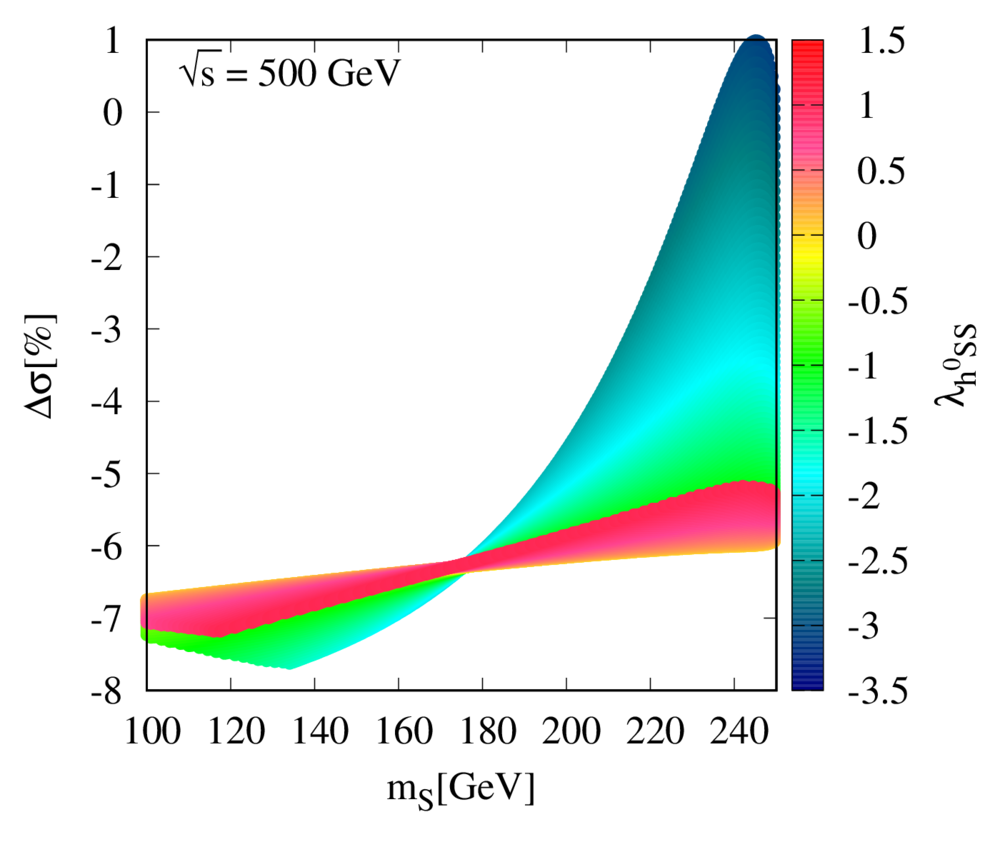}
	\includegraphics[width=0.31\textwidth]{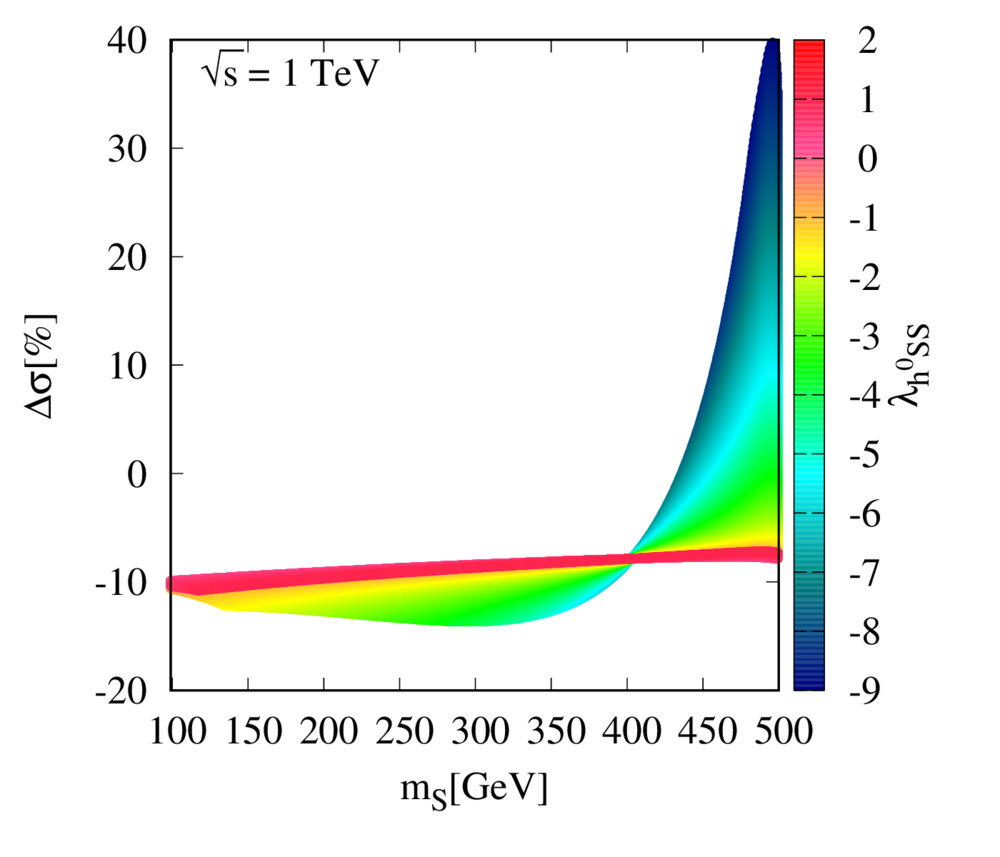}\\
	\includegraphics[width=0.31\textwidth]{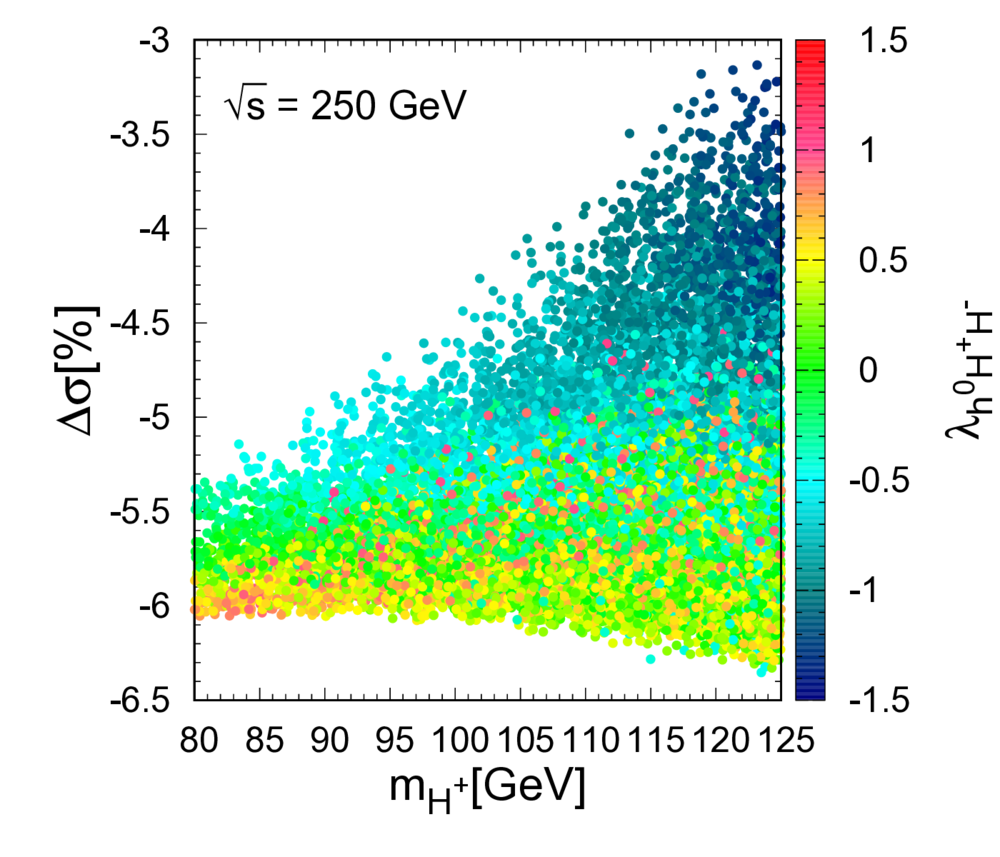}\includegraphics[width=0.31\textwidth]{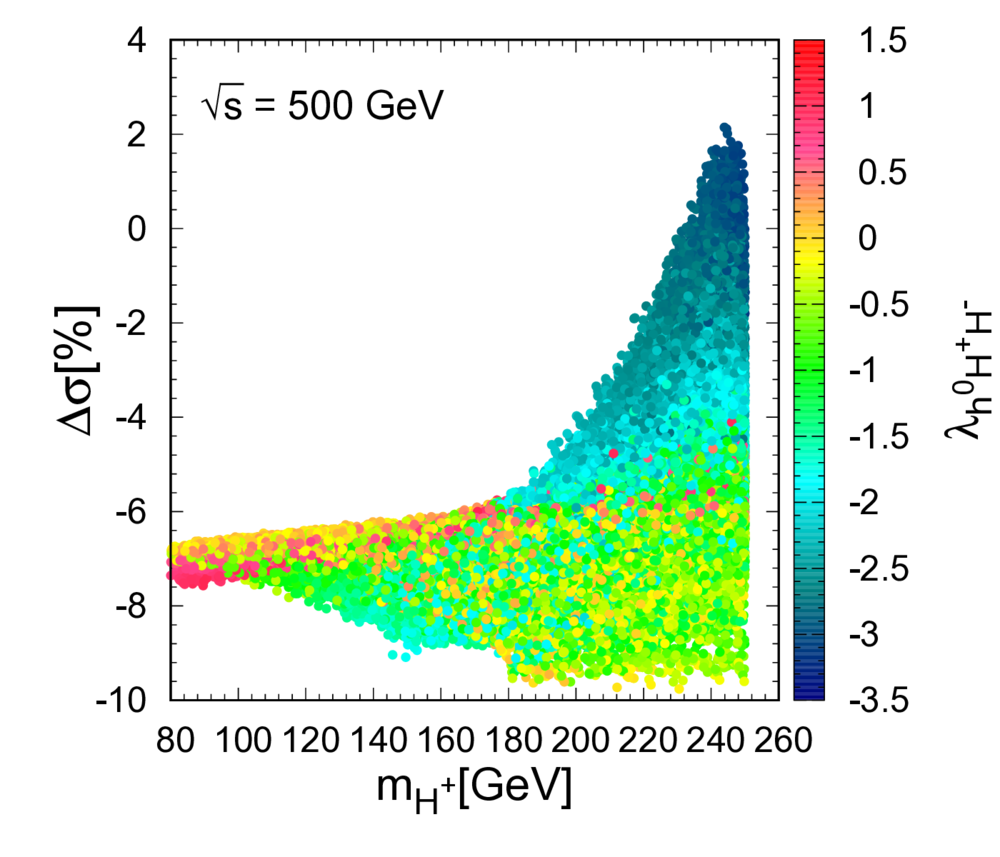}
	\includegraphics[width=0.31\textwidth]{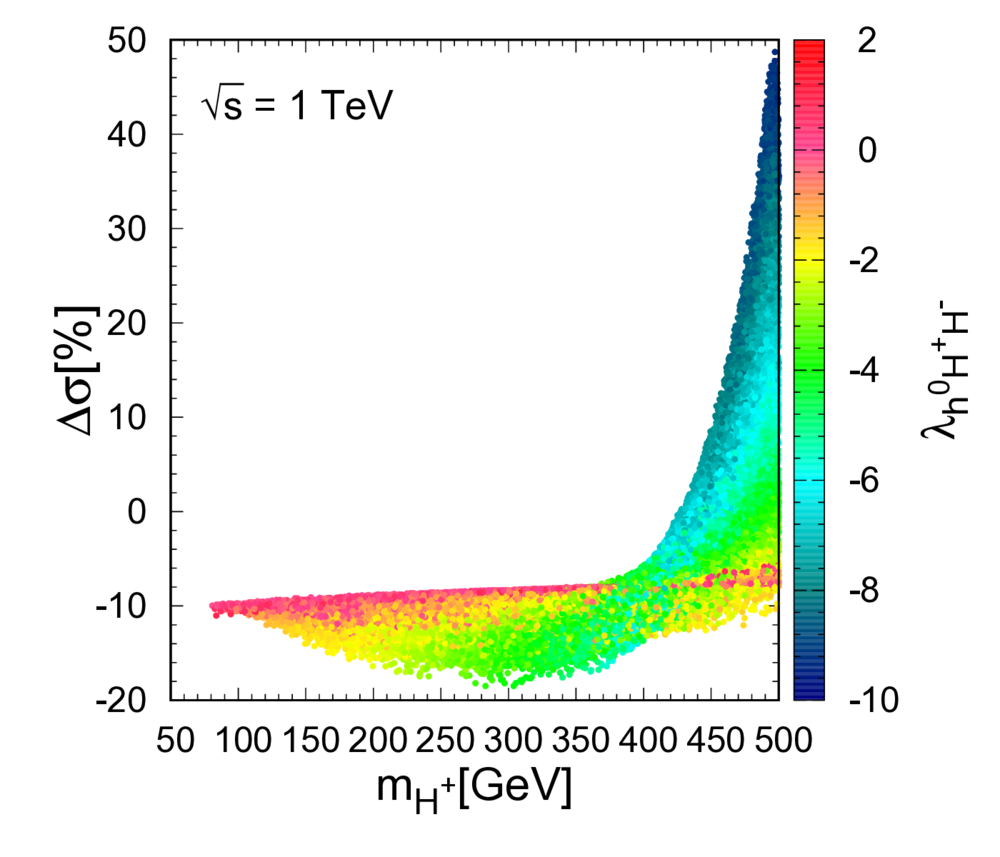}\\
	\includegraphics[width=0.31\textwidth]{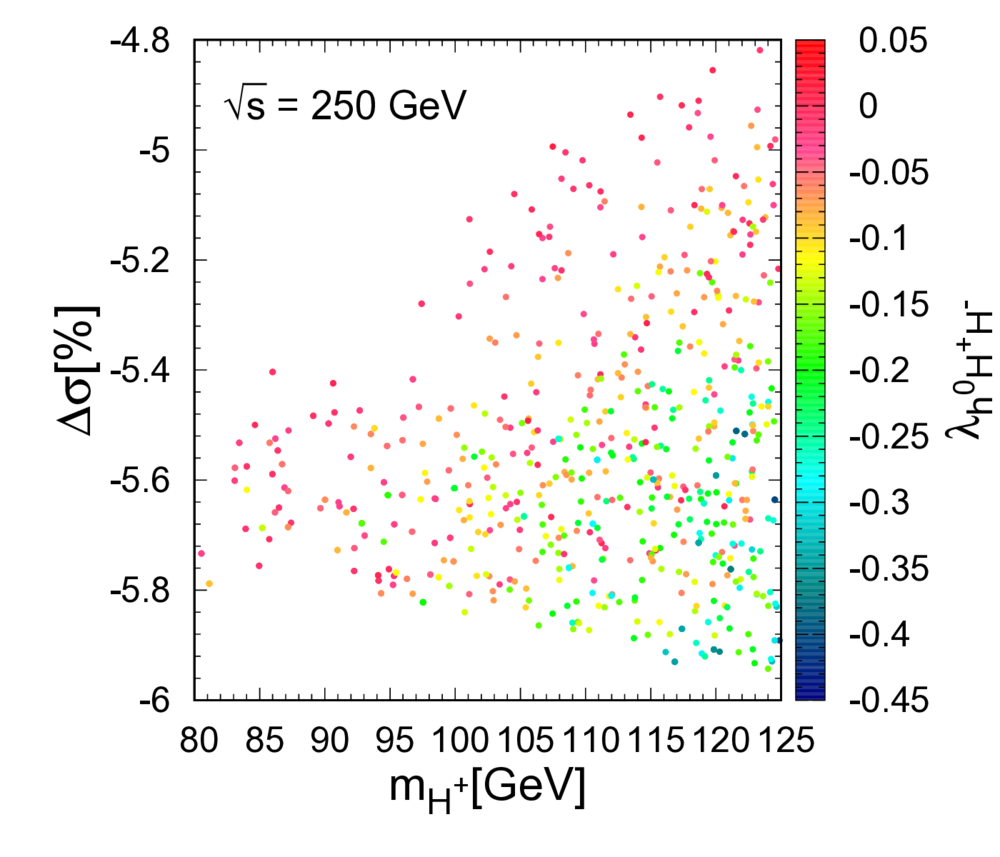}\includegraphics[width=0.31\textwidth]{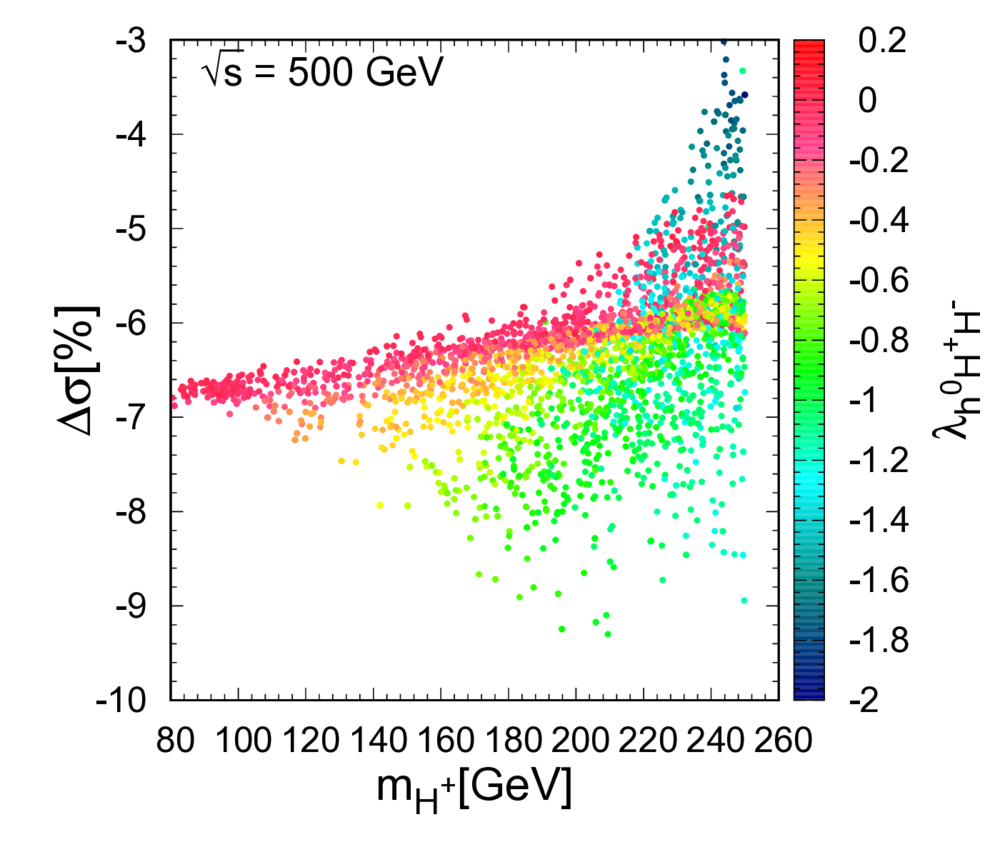}
	\includegraphics[width=0.31\textwidth]{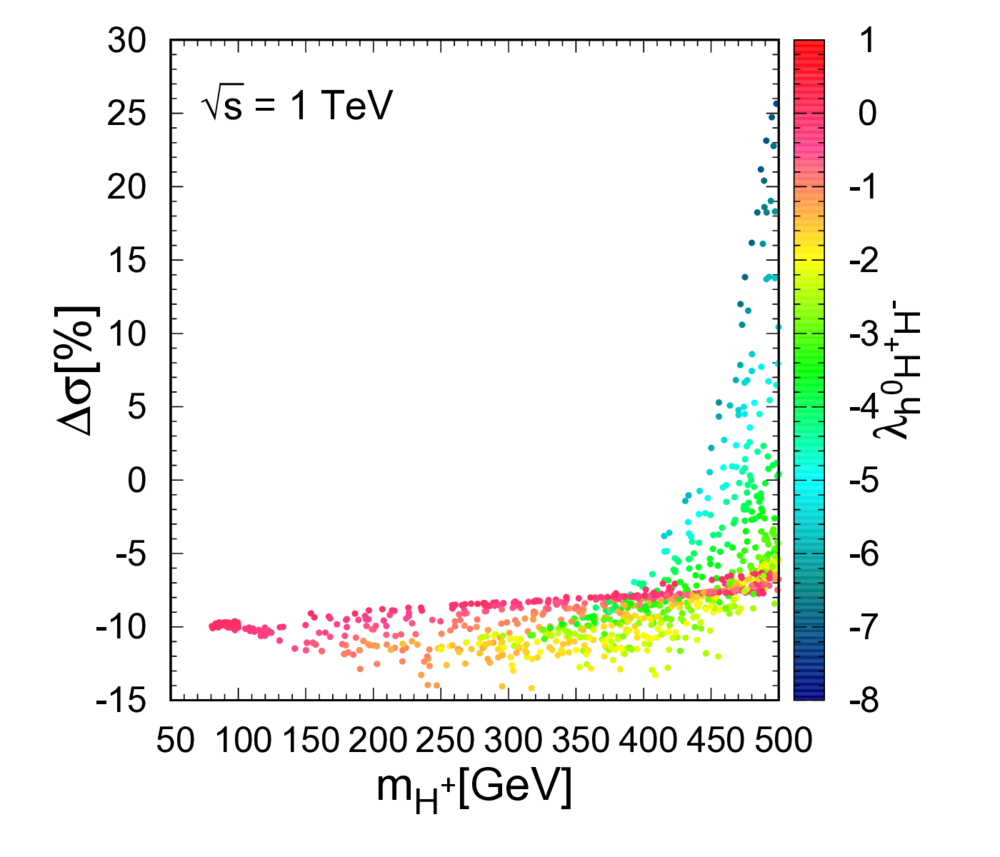}
	\caption{Electroweak corrections to  $e^+ e^- \to H^+ H^-$ for $\sqrt{s}=250$ GeV, 500 GeV, and 1000 GeV  as a function of $m_S$  and the {$\lambda_{hSS}$ coupling} normalized to the VEV of the SM Higgs are shown. Plots in the upper panel show scenario I, and the middle and lower panels show the nondegenerate scenario before (scenario II) and after applying dark matter constraints (scenario III), respectively. }
	\label{fig:eeHH}
\end{figure}

With these crucial lessons on weak corrections learned from scenario I, we are ready to further explore the whole parameter spaces of the IDM and illustrate the behavior of weak corrections for all three scenarios, as shown in Fig.~\ref{fig:eeHH}.

The upper panel of Fig.~\ref{fig:eeHH} is devoted to scenario I, where the ratio of EW corrections to $e^+e^- \to H^+ H^-$ as a function of the inert scalar masses $m_S$ are shown, while the $\lambda_{h^0SS}/v=-\lambda_3$ coupling  is labeled as a color bar. For the c.m. energy $\sqrt{s}=250$ GeV, the parameter $m_S$ can be measured in the range $[80,125]$ GeV, and the ratio $\Delta\sigma$ can change in the range $[-5.9\%, -3.8\%]$;  for the c.m. energy $\sqrt{s}=500$ GeV, {{the parameter $m_S$ is kinematically accessible up to $250$ GeV}} and $\Delta\sigma$ can change in the range $[-7.8\%, 1\%]$. For these two cases, the ratio of EW corrections is not very significant, which is due to the fact that the range of $\lambda_3$ is limited to $[-3.5, 1.5]$. In contrast, for the c.m. energy $\sqrt{s}=1$ TeV,  {{the parameter $m_S$ is kinematically accessible up to $500$ GeV}}, which in turn implies a larger range for $\lambda_3\in [-9,2]$. Then the ratio of EW corrections can be in the range $[-13\%, 38\%]$. In this case, the relative corrections become substantial for $m_S$ close to the threshold production, where 
the absolute values of $\lambda_3$ can be quite large (say $|\lambda_3|\sim 9$).
Consequently, 
the relative size of loop corrections increases and can reach $40\% $.


For scenario II, the theoretical parameters $\lambda_{4,5}$ are nonvanishing and could contribute to the vacuum stability  and unitarity constraints. Therefore, when compared with scenario I, two more dimensions are added to the theoretical parameter space. Similar to scenario I, the parameter $\lambda_3$ should be constrained by the Higgs data, like the branching fraction measurement of  $h^0\rightarrow \gamma\gamma$, since a light charged scalar can have a significant contribution to this branching fraction. In the case with $\sqrt{s}=250$ GeV, as shown by the plots of the middle row of Fig.~\ref{fig:eeHH}, due to kinematics, the range of charged scalar mass that the machine can measure is $[80, 125]$ GeV.  It is found that the diphoton signal strength puts a severe constraint on the parameter $\lambda_3$, which can only be within the range $[-1,1]$. Thus, the effect of {the $h^0H^+H^-$ coupling} is as small as the scenario I where the allowed ratio of radiative corrections can spread from $-6.5\%$ to $ -3\%$. At $\sqrt{s}=500$ GeV, the allowed parameter $\lambda_3$ can be in the range $[-1.5, 3.5]$ and the radiative corrections start from $ -10\%$ and can go up to $ 2\%$ near the end point region of $m_{H^\pm}$. In contrast, in the case with the c.m.~energy  $\sqrt{s}=1$ TeV, the allowed range of $\lambda_3$ becomes wider, it can change from $-10$ to $2$, which leads to a significant enhancement for the ratio of EW corrections as shown in the plot. {The  $h^0H^+H^-$ coupling} can contribute through the charged scalar wave function renormalization and the triangle diagrams.  The ratio of EW corrections is large and can even go up to $ 50\%$ near the end point (for a charged scalar mass around $m_{H^+}=500$ GeV).

For scenario III, we show the ratio of EW corrections in the lower panel of Fig.~\ref{fig:eeHH}, where we have taken into account all existing constraints on the parameter space of the IDM, especially the dark matter relic density, direct search constraints, and collider search of dark matter candidates. We can see that at  $\sqrt{s}=250$ GeV the dark matter constraints shrink the allowed range of $\lambda_3$ and the ratio of EW corrections can only vary between $-6\%$ and $-4.8\%$.
 At  $\sqrt{s}=500$ GeV, the ratio of EW corrections can only be negative, and it can only change from $-9\%$ to $-3\%$. At $\sqrt{s}=1$ TeV, a larger theoretical parameter space can be probed. Accordingly, a wider range of $\lambda_3$ leads to  a larger ratio of EW corrections, which can spread from $-15\%$ to $30\%$ near the end point region.

When the allowed points in the parameter space are compared for scenarios II and III, it is noteworthy that the dark matter constraints can kill almost $99\%$ points, as demonstrated in the lowest panels of Fig. \ref{fig:eeHH}, only sporadic points are allowed for scenario III.

It should be pointed out that the ratio of EW corrections could be large and can reach $20\%$ or higher near the end point region, like in the $\sqrt{s}=1$ TeV case, due to the small cross section, the production rate might also be small. For example, in the IDM5 case, when $m_{H^\pm} = 465$ GeV, the cross section is 2 fb or so. Nonetheless, when the integrated luminosity is large enough (in the case of 2/ab, we can have 4000 signal events), there is a chance to examine this loop induced effect if we could know the model parameters precisely from other measurements. 
Finally, we would like to point out that the EW corrections are rather significant for high 
c.m.~energy $\sqrt{s}=1$ TeV than they are for low energy cases  $\sqrt{s}=250$ TeV or 500 GeV. This is 
 because the contributions from the boxes are rather important for high energy.

\section{Benchmark points}
\label{sec:BPs}
\begin{table}[!htb]
\renewcommand\arraystretch{1.2}
\centering
\begin{tabular}{|c|rrrrrr|}
\hline\hline
BP & BP1  & BP2 &  BP3 &  BP4& BP5  &BP6\\
\hline
$m_{H^\pm}$ (GeV)     &  116.8&    123.4&   209.5 &   243.7 &     295.4 & 472.9 \\\hline
$m_{H^0}$ (GeV)         &   57.0&     121.9&   122.9 &    59.3  &     204.1 & 181.4  \\\hline
$m_{A^0}$  (GeV)        &  102.3&    200.0&   125.2 &   238.3 &     205.7 & 473.5 \\\hline
$\mu_2^2$  (GeV$^2$)&3159.5&14723.8&15037.5&  3558.6 &  41195.9&32220.8 
\\\hline
 $\lambda_{L} (10^{-3})$                  &  1.537 & 2.416&1.151& -0.798& 7.847&11.819  \\\hline
 $\lambda_{S}$                                &   0.124& 0.427& 0.011&  0.900& 0.019& 3.246\\  
  \hline
 $\Omega h^2 (\times 10^{-2})$       &10.028           &                0.340&             0.203  &         5.428   &              0.185 &0.028 \\\hline
 $\Gamma_{H^\pm}$ (GeV) & $5.87\times10^{-4}$ & $5.48\times 10^{-12}$ & $6.77\times 10^{-2}$ &$2.58$ &$2.18\times 10^{-1}$ & $1.86\times 10^1$ \\\hline
  $\Gamma_{A^0}$ (GeV)    & $1.06\times10^{-4}$ &  $1.23 \times 10^{-2}$ & $5.19 \times 10^{-11}$ & $2.00$ & $8.85 \times 10^{-12}$ &$1.79\times 10^1$ \\\hline
$Br(h^0 \to H^0H^0)$                       &$0.47\%$       &                   -   &                       - &      $0.10\%$&                      - &-\\\hline
$Br(A^0 \to W^{\pm(*)} H^{\mp})$    &             -        & $\sim 76.5\%$&                       - &                    -&                       -&$\sim 0\%$ \\\hline
$Br(A^0 \to Z^{(*)} H^0)$                  &$100\%$        & $\sim 23.5\%$&         $100\%$ &        $100\%$& $100\%$         &$\sim 100\%$\\\hline
$Br(H^\pm \to W^{\pm(*)} A^0)$       &$\sim 0\%$    &                       -& $\sim 41.6\%$ &    $\sim 0\%$& $\sim 43.8\%$& -\\\hline
$Br(H^\pm \to W^{\pm(*)} H^0)$      &$\sim 100\%$ &          $100\%$& $\sim 58.4\%$ &$\sim 100\%$& $\sim 56.2\%$& $100\%$\\
\hline\hline
\end{tabular}
\caption{Benchmark points consistent with collider experiments  and dark matter constraints on the relic density  are proposed. Decay information of $H^0$, $A^0$, and $H^\pm$ are also given.}
\label{tab:BPs}
\end{table}

In Table~\ref{tab:BPs} we propose six benchmark points from scenario III that are consistent with current collider experiments and dark matter searches. It is important to mention that several constraints from current long-lived particles searches on the IDM exist in the literature \cite{Heisig:2018kfq,Belyaev:2020wok}. However, we did not take them into account in the present analysis. But  as it has been discussed in \cite{Kalinowski:2020rmb}, limits from quasistable charged particle searches can be evaded if we set an upper bound on the charged scalar lifetime of $\tau \le 10^{-7}$s, this implies a lower bound on the total decay width of the charged scalar of $\Gamma_{H^+}\ge 6.582\times 10^{-18}$ GeV, which is respected by our BPs.

 It should be mentioned that we assume that IDM alone is not sufficient to accommodate the whole dark matter content of the Universe, therefore, we only demand that the relic density of IDM is smaller than the one required by experiments. As a matter of fact, there might be other extra sectors {that} can contribute to the relic density of the Universe but have not {been} taken into account in this work.  One example is the right-handed neutrino sector, another possible candidate {for} dark matter are {the} axion or axionlike particles, and there could be others as well. Coupling constants and decay information are also presented in the table. In order to examine the radiative effects of weak interactions, we have deliberately proposed three BPs, i.e. BP2, BP4, and BP6, which are close to the threshold region of certain c.m.~energies of 250 GeV, 500 GeV, and 1 TeV, respectively.

There are a few comments on the features of these benchmark points. (1) According to our previous findings given in \cite{Abouabid:2020eik}, it is found that both $A^0$ and $H^0$ can be the dark matter candidates. For the sake of simplicity, in these BPs, we assume that the dark matter candidate is $H^0$. (2) For BP2, the charged scalar has a long life time, which can trigger the signature of a displaced vertex caused by a massive charged particle. For BP3 and BP5, the $A^0$ has a long life time, which can trigger the signature  of a displaced vertex caused by a neutral heavy particle. While for the rest of BPs, both $A^0$ and $H^\pm$ can promptly decay when produced. (3) For BP1 and BP4, the extra invisible decay mode $h^0 \to H^0 H^0$ can be open, which will lead to a larger invisible decay branching for the SM-like Higgs boson. According to the CDR of CEPC \cite{CEPCStudyGroup:2018ghi} and a recent Monte Carlo study \cite{Tan:2020ufz}, the future Higgs factory of CEPC with $\sqrt{s}=250$ GeV could have the potential to determine the branching fraction of the invisible decay of $h^0$ down to $0.3\%$. Therefore, it is possible to probe BP1 at the CEPC with $\sqrt{s}=250$ GeV not only through $e^- e^+ \to H^+ H^-$ but also via the process $e^- e^+ \to Z h^0$.

Another comment is on the decay widths of charged scalars. When decay widths are tiny, the narrow width approximation is appropriate, as benchmark points 1$-$4 and 5 are narrow enough since decay widths are only $1\%$ of masses. For the case where decay widths are not tiny, as benchmark point 6 demonstrated, where decay width is around $4\%$ of its mass, then the narrow width approximation might not be precise enough to describe the production $H^+ H^-$ and the consequent decay products. A better method to describe $e^+ e^- \to H^+ H^- \to W^+ W^- H^0 H^0$ is to include the decay widths of $H^\pm$ in the matrix elements, instead of producing $e^+ e^- \to H^+ H^-$ and then decaying $H^\pm$ to $W^\pm H^0$. Nonetheless, when its decay width is $10\%$ smaller than its mass, our numerical calculation demonstrate, the difference is acceptable. Only when its decay width is too big, say larger than $20\%$, the cross section in this case should be computed carefully.\\
Furthermore, we have explored the dependence of the decay width on the charged scalar mass, and we have found that the total decay width of the charged scalar over its mass in  scenario III cannot exceed 5\%.

In Table~\ref{Tab:XSBPs}, weak corrections, QED corrections, the LO and full one-loop cross sections of these BPs are provided. Generally speaking, at $\sqrt{s}=250$ GeV and $\sqrt{s}=500$ GeV, {{the weak corrections are negative since the size (absolute value) of the $\lambda_{h^0H^+H^-}$ coupling is small (say from -1.5 to 1.5), as shown in Fig.~\ref{fig:eeHH}. Only when the charged scalar is heavy enough (say 450 to 500 GeV) and the absolute value of the $\lambda_{h^0H^+H^-}$ coupling can be large (say from -10 to 2 for scenario II and  from -8 to 1 for scenario III),}} the weak corrections $\Delta_{\mathrm{weak}}$ can be positive when the masses of charged scalar are close to the threshold region when $\sqrt{s}=1000$ GeV, as demonstrated by BP6. {{It should be mentioned that the positive contribution comes from diagrams with charged scalars in the loop.}}

\begin{table}[!htb]
\renewcommand\arraystretch{1.2}
\centering
\begin{tabular}{|c|crrrr|}
\hline\hline
  \multicolumn{1}{|c|}{$\sqrt{s}$ (GeV) } & \multicolumn{1}{c}{BP}
&\multicolumn{1}{c}{$\sigma^{0}$ (fb)}&\multicolumn{1}{c}{$\Delta_{\mathrm{weak}}$(\%)}
&\multicolumn{1}{c}{$\Delta_{\mathrm{QED}}$(\%)}&\multicolumn{1}{c|}{$\sigma^{\mathrm{NLO}}$ (fb)} \\
\hline
\multirow{2}*{250}
&BP1&23.940         &-5.941         &-0.138           &22.484  \\
&BP2&2.300          &-4.825         &-0.475            &2.178  \\
\hline
\multirow{4}*{500}
&BP1&86.733  &-7.267  & 0.261  &80.655  \\
&BP2&82.604  &-6.373  & 0.247  &77.543  \\
&BP3&20.593  &-9.213  & 0.033  &18.703   \\
&BP4& 1.446  &-2.842  &-0.332  &1.400 \\
\hline
\multirow{6}*{1000}
&BP1&28.563 &-10.631 & 0.421 &25.647  \\
&BP2&28.280 & -9.360  & 0.409 &25.749  \\
&BP3&23.284 &-13.896 & 0.286 &20.115 \\
&BP4&20.715 &-10.987 & 0.249 &18.491 \\
&BP5&16.365 & -14.023 &0.194 &14.102 \\
&BP6& 1.092& 11.543 &-0.174 &1.216 \\
\hline\hline
\end{tabular}
\caption{Weak corrections, QED corrections, the LO and full one-loop cross sections of BPs are provided.}
\label{Tab:XSBPs}
\end{table}

From the above table, one can see that at the $e^+e^-$ colliders, in BP1, BP2, BP4 and BP6,  $e^+e^- \to H^+ H^-$ would lead to $W^+W^-H^0H^0$ final state, which in turn could lead to a final state with dileptons and missing energy. While for BP3 and BP5 one could have the following final states: $H^+ H^-\to W^+W^-H^0H^0$  or  $H^+ H^-\to W^+W^-H^0A^0\to W^+W^-Z^0H^0H^0$, which could give  dilepton events as well as multileptons. 

At the LHC, the situation is slightly different because we have the following production mechanisms for charged scalar: $pp\to H^+ H^-$ ,  $pp\to H^+ H^0$,  and 
 $pp\to H^+ A^0$. For the charged scalar pair production, similar final states as for $e^+e^-\to H^+H^-$ can be obtained.
 While the process $pp\to H^+ H^0$ (respectively, $pp\to H^+ A^0$ ) 
 would give $W^+H^0H^0$ (respectively, $W^+H^0A^0\to W^+Z^0H^0H^0 $),  or 
  $W^+A^0H^0\to W^+Z^0 H^0H^0$  (respectively, $W^+A^0A^0\to W^+Z^0Z^0 H^0H^0$ ), which leads to final states with single/multileptons and missing energy. 
At the LHC, it should be pointed out that in order to pinpoint the electroweak corrections, the 
QCD corrections should first be reliably calculated, which are typically quite large.
 
{Moreover, there are two more comments on our BPs: (1) We have checked our BPs in Table~\ref{tab:BPs}  and confirmed that they can survive even in light of the new direct detection results from the LUX-ZEPLIN Experiment~\cite{LZ:2022ufs}.
(2) For multicomponent dark matter situations, a scaling factor is included in Ref.~\cite{Ilnicka:2015jba}; however, this factor is not taken into account in our work. Such a factor will ease the restrictions from direct search because we require that the relic density from IDM is not greater than the one from $Planck$ data.
Based on the fact that our BPs have already passed all the constraints in our analysis, it  is easy to conclude that they will all survive if the factor is included.
}

\section{Conclusions and discussions}
IDM is a simple model that can solve the problem of dark matter. Such a model predicts a charged scalar in its spectrum. The observation of such charged scalar  
either at LHC or at the future $e^+e^-$ colliders would be a conclusive evidence of physics beyond SM.

Future $e^+e^-$ colliders would provide an opportunity to  measure the charged scalar cross section and its properties precisely. Following the standard renormalization scheme of the SM,  
in this work, we study the radiative corrections of the new  physics process $e^- e^+ \to H^+ H^-$ in the IDM.
The dimensional regularization is used to evaluate the one-loop Feynman amplitudes in the Feynman$-$'t Hooft gauge. 
We employed comprehensive on-shell scheme renormalization, which means that not only the particle masses and fields, but also the coupling constant, were renormalized using on-shell conditions.
As a consequence, our predictions are totally independent of the renormalization scale $\mu_r$. 
Nevertheless, a new {scale} $m_Z$ is introduced via Eq.~(\ref{eqn:renor}) where the large logarithms from light fermions are absorbed into the redefinition of the running coupling constant. The running coupling constant is changed from 1/137.036 to 1/128.943 correspondingly. This yields a $13\%$ difference in the LO predictions. 
But this new scale dependence is greatly reduced at the NLO. 

We have considered the QED corrections and checked that the IR divergences cancel when we add 
the virtual and real photon emissions. We have also used the resummed cross section to cure the well known Coulomb singularity.
In addition, collinear divergences in our calculation appear as terms proportional to $\log(m_e)$. After including the counter term from the structure function of an electron, such divergent terms should vanish in the final result. In order to check this, we vary the mass of the electron {by}  a factor of $k$ from $2^{-6}$ to $2^6$, namely $m_e$ is taken $k\times 0.511$ MeV. It is found that the result remains unchanged when $k$ varies. 

We have examined the size of the weak corrections for three representative c.m.~energies: $\sqrt{s}=250$, $\sqrt{s}=500$, and $\sqrt{s}=1000$ GeV. 
 After taking into account the theoretical constraints and experimental bounds for the scenario III, we have found that the weak corrections are still sizable; i.e.
 the weak corrections are around $-6\% \sim-5\%$ for $\sqrt{s}=250$ GeV, $-10\% \sim -3\%$ for $\sqrt{s}=500$ GeV, and $-15\% \sim +25\%$ for $\sqrt{s}=1000$ GeV, 
 as shown in Fig.~\ref{fig:eeHH}. The origin of those sizable corrections {is the $h^0H^+H^-$ coupling} which could become large for some  configuration of the parameters. We have shown that  the size of the radiative corrections, are typically of the order 5$-$25\%, which makes their proper inclusion in any phenomenological studies and analyses for $e^+e^-$ colliders indispensable. 

 From those allowed parameter points, we have proposed six benchmark points as given in Table~\ref{tab:BPs} for future lepton collider searches.   We provide the parameters, the total decay width, as well as the branching fractions of the neutral and charged scalars.  
According to those BPs,  we have also discussed the signature of  $e^+e^- \to H^+ H^-$  followed  by the charged scalar decays and show that it would lead to a final state with multileptons and missing energy. 
 At the LHC, one can get similar signatures through  $pp \to H^\pm H^\mp ;  H^\pm A^0 ;  H^\pm H^0$ production.


It was noteworthy that initial state radiation can reduce the production rate of the process $e^+ e^- \to Z h^0$ for $\sqrt{s}=240/250$ GeV by a factor of $10\%$ or so \cite{Mo:2015mza}, according to the simulation done with the public code \texttt{WHIZARD} \cite{Kilian:2007gr}. For the process $e^+ e^- \to H^+ H^-$, the NLO cross section $\sigma^1$ satisfies $\sigma^1 = |M^0 + M^1|^2 \approx |M^0|^2 + 2 \mathrm{Re} (M^0 M^{1*}) $, hence the initial state and final state radiations are simply linearly summed in the matrix element of $ M^1$. Therefore, at the $O(\alpha)$ order, the initial state radiation of the process $e^+ e^- \to H^+ H^-$ is just the same as that of the process $e^+ e^- \to Z h^0$, i.e., such a reduction in the production rate also holds for the process $e^+ e^- \to H^+ H^-$ for $\sqrt{s}=240/250$ GeV. The reduction can be around $-10\%$ in the $M_Z$ scheme, as shown in Ref. \cite{Xie:2018yiv} for $\sqrt{s}=240$ GeV. While for the cases $\sqrt{s}=500$ GeV and $\sqrt{s}=1$ TeV, due to the effects of radiative return \cite{Chakrabarty:2014pja,Karliner:2015tga,Greco:2016izi}, it is expected that the initial state radiation can enhance the cross section of $e^+ e^- \to H^+ H^-$ when $m_{H^\pm}$ is small, as in the cases of BP1 and BP2.

Meanwhile, at the $O(\alpha)$ order, the initial state radiation and final state radiation can be treated individually. In order to cure the Coulomb singularity arising from the final state interaction, the final state radiations can be partially resumed, and the cross section can be modified as given in Eq.~(\ref{eqn:resum1}). In terms of our numerical results given in Table~\ref{Tab:resum}, the Coulomb singularity can be reliably removed. And the resummation effect is represented by the factor $|\phi(0)|^2$. Thus the final state radiation can be correctly evaluated, which can increase the total cross section by a factor  from $1\%$ to $49\%$. It reaches its maximum when the invariant mass of the pair of charged scalar approaches the total collision energy, as shown in Table~\ref{Tab:resum} for $\sqrt{s}=500$ GeV. 

For higher order $O(\alpha)$ corrections, there exist interference terms between initial state radiation and final state radiation, appropriate treatment of these two radiations is beyond the scope of current work.

\label{sec:conclusions}
\section*{Acknowledgements}
We thank Jianxiong Wang for helpful discussions about FDC program. Q. S. Y. is supported 
by the Natural Science Foundation of China under the grant No. 11475180 and 
No. 11875260. B. G. is supported by the Natural Science Foundation of China Grants No.~11475183, No.~11975242, and No.~12135013. J. E. F.  would like to thank the HECAP section of the Abdus Salam International Centre for Theoretical Physics (ICTP)  for hospitality and financial support where part of this work has been done.
The work of H. A, A. A, and J. E. F.  is supported by the Moroccan Ministry of Higher 
Education and Scientific Research MESRSFC and CNRST: Projet PPR/2015/6.

\appendix
\section{Details about the renormalization of charge}
\label{sec:charge}
As mentioned in the main text, we have used an ``on-shell" renormalization scheme not only for the masses and fields of particles but also for the charge. 
By adopting this approach, our results are independent of the renormalization scale $\mu_r$.
However, in order to resum the large logarithms arising from light fermions vacuum polarization, a new scale $Q=m_Z$ is introduced. {Further details are presented below.} 

{ The electric charge renormalization is carried out} following Refs.~\cite{Denner:1991kt,Denner:2019vbn}. 
The renormalization constant is obtained in the Thomson limit with the condition { by imposing the condition
}
\begin{equation}
\bar{u}(p)\hat{\Gamma}^{ee\gamma}_{\mu}(p,p)u(p)|_{p^2=m_e^2}=ie\bar{u}(p)\gamma_\mu u(p), 
\end{equation}
which gives
\begin{equation}
\delta Z_e (0)=-\dfrac{1}{2}\delta Z_{AA}-\dfrac{s_W}{c_W}\dfrac{1}{2}\delta Z_{ZA}
=\dfrac{1}{2}\Pi(0)-\dfrac{s_W}{c_W}\dfrac{\sum^{AZ}_T(0)}{m_Z^2}
\label{eqn:ze0}
\end{equation}
where the renormalization constants are defined as follows (bare quantities are denoted by a subscript ``0"):
\begin{equation}
\begin{aligned}
{e_0\mu^{\varepsilon}}&{=Z_e e\mu_r^{\varepsilon}=(1+\delta Z_e)e\mu_r^{\varepsilon}}, \\[3mm]
\left(\begin{array}{c} Z \\ A \end{array}\right)_0
&=
\left(
\begin{array}{cc}
  1+  \frac{1}{2}\delta Z_{ZZ}  &  \frac{1}{2}\delta Z_{ZA}  \\  \frac{1}{2}\delta Z_{AZ} &     1+  \frac{1}{2}\delta Z_{AA}
\end{array}
\right)
\left(\begin{array}{c} Z \\ A \end{array}\right).
\label{eqn:e0}
\end{aligned}
\end{equation}
{Here $\varepsilon\equiv (4-D)/2$ with $D$ being the space-time dimension, $\mu$ is the arbitrary mass parameter introduced for bare charge, $\mu_r$ is the renormalization scale of charge, $e_0$ is the bare dimensionless charge and $e$ is the renormalized one.}
For the remaining quantities in Eq.~(\ref{eqn:ze0}),  ``$\Pi$" is defined as
\begin{equation}
\Pi(k^2)\equiv\dfrac{\sum^{AA}_T(k^2)}{k^2}, \quad
\Pi(0)=\lim_{k^2\rightarrow 0}\dfrac{\sum^{AA}_T(k^2)}{k^2}=\dfrac{\partial\sum^{AA}_T(k^2)}{\partial k^2}\biggr|_{k^2=0}, 
\end{equation}
and $\sum^{AA(AZ)}_T$ {denotes the transverse part of the $AA(AZ)$ self-energy.}
This is just the ``zero-momentum" scheme in Ref.~\cite{Degrande:2014vpa}.

It should be noted that in the calculation of $\Pi(0)$, nonperturbative strong interaction effects cannot be neglected. In order to treat them properly, we first introduce a universal quantity:
\begin{equation}
\Delta \alpha({Q})\equiv\Pi(0)-\mathrm{Re}\Pi({Q^2})
\label{eqn:dalpa}
\end{equation}
to denote the difference of $\Pi$s when the external photon is on shell and off shell, respectively.
Here {$Q$} satisfies {$Q^2=k^2$} where $k$ is the momentum of an off-shell external photon.
At the one-loop level, it is easy to separate this $\Delta \alpha$ into several parts according to the particles inside the photon's self energy, i.e.
\begin{equation}
\begin{aligned}
\Delta \alpha=&(\Delta \alpha)_{f}+(\Delta \alpha)_{b}
\end{aligned}
\end{equation}
where the labels  ``$f$" and ``$b$" are used to denote the contributions from fermions and bosons. For the convenience of discussion below, the $(\Delta \alpha)_{f}$ part can further be expressed as 
\begin{equation}
\begin{aligned}
(\Delta \alpha)_{f}=&(\Delta \alpha)_{\mathrm{\ell}}+(\Delta \alpha)_{\mathrm{q}}^{(5)}+(\Delta \alpha)_{\mathrm{top}}
\label{eqn:ddalpha}
\end{aligned}
\end{equation}
and the labels  ``$\ell$", ``q", and ``top" are used to denote the contributions from leptons, light quarks and top quarks, respectively. Here ``5" in $(\Delta \alpha)_{\mathrm{q}}^{(5)}$ refers to the number of quark flavors. $\Pi$ can also be separated in the same way, so we will also apply the labels to $\Pi$.

Since we are trying to treat the nonperturbative effects in $\Pi(0)$, we focus on the ``q" part here, which can be expressed as 
\begin{equation}
(\Delta \alpha)^{(5)}_{\mathrm{q}}({Q})=\Pi^{(5)}_{\mathrm{q}}(0)-\mathrm{Re}\Pi^{(5)}_{\mathrm{q}}({Q^2}). 
\label{quarkal}
\end{equation}
Although $\Pi^{(5)}_{\mathrm{q}}(0)$ can be calculated perturbatively, it is unreliable due to its nonperturbative nature at low energy regions. Hence we try to exact it from data. At the scale {$Q=m_Z$}, the lhs of Eq.~(\ref{quarkal}) can be obtained from experiment measurements using the dispersion relation, i.e.
\begin{equation}
(\Delta \alpha)^{(5)}_{\mathrm{q}}(m_Z)=(\Delta \alpha)^{(5)}_{\mathrm{hadron}}(m_Z). 
\end{equation}
Here $(\Delta \alpha)^{(5)}_{\mathrm{hadron}}(m_Z)$ is the experimental data and its value can be found from PDG~\cite{Tanabashi:2018oca} as $\Delta\alpha^{(5)}_{\mathrm{hadron}}(m_Z)=0.02764$. Meanwhile at this scale $\Pi_{\mathrm{q}}^{(5)}(m^2_Z)$, the second term on the rhs of Eq.~(\ref{quarkal}), can be reliably computed by the perturbation expansion. Combing these with Eq.~(\ref{quarkal}), we have
\begin{equation}
\Pi_{\mathrm{q}}^{(5)}(0) =\mathrm{Re}\Pi^{(5)}_{\mathrm{q}}(m_Z^2)+\Delta\alpha^{(5)}_{\mathrm{hadron}}(m_Z),
\label{eqn:pi0non}
\end{equation}
in which the nonperturbative effects have been properly included. Then $\Pi(0)$ is obtained as
\begin{equation}
\Pi(0)= \Pi_{\mathrm{q}}^{(5)}(0) +\Pi_{\mathrm{rest}}(0) =
\mathrm{Re}\Pi^{(5)}_{\mathrm{q}}(m_Z^2)+\Delta\alpha^{(5)}_{\mathrm{hadron}}(m_Z)+\Pi_{\mathrm{rest}}(0).
\label{eqn:pi}
\end{equation}
Here ``rest" denotes the contribution of ``$\ell$", ``top", and ``$b$" parts. 

Once $\Pi(0)$ is determined by  Eq.~(\ref{eqn:pi}), the renormalization of charge defined by Eq.~(\ref{eqn:ze0}) corresponds to the electromagnetic fine-structure constant obtained in the Thomson limit: $\alpha(0)=1/137.036$.

As pointed out in Sec. 8.2.1 of Ref.~\cite{Denner:1991kt} , {the }above renormalization of charge [labeled as $\alpha(0)$ scheme for convenience] {is carried out} at zero momentum transfer, where the relevant scale is set by the masses of fermions. 
They are much smaller than the relevant scales in high energy experiments. 
Hence, when working on processes in high energy experiments, such as this work, the large ratio of these different scales will cause large logarithms. 
{
These large logarithms can be summarized in the universal quantity introduced above, $\Delta \alpha({Q})$.
}

The leading logarithms in  $\Delta\alpha$  arise from fermionic contributions and can be correctly resummed to all orders in perturbative theory by the replacement: 
\begin{equation}
1+(\Delta\alpha)_{\mathrm{LL}}\rightarrow \dfrac{1}{1-(\Delta\alpha)_{\mathrm{LL}}}, 
\end{equation}
where a label ``LL" is used to denotes the leading logarithms. 

Meanwhile, since not only the leading logarithms but  also the whole fermionic contributions are gauge invariant, we can resum the latter, i.e. 
\begin{equation}
1+\Delta\alpha =1+(\Delta\alpha)_{f}+ (\Delta\alpha)_{b}\rightarrow \dfrac{1}{1-(\Delta\alpha)_f}+ (\Delta\alpha)_{b} .
\end{equation}

On the other hand, the large logarithms in $\Delta\alpha$ originate from  the renormalization constant $\delta Z_e(0)$ given in Eq.~(\ref{eqn:ze0}); hence they will appear wherever $\alpha$ appears in lowest order and can be taken into account by replacing the lowest order $\alpha$ by a running $\alpha({Q})$ as 
{
\begin{equation}
\alpha_{\mathrm{lowest}}=\alpha(0) \rightarrow \alpha({Q})=\dfrac{\alpha(0)}{1-(\Delta\alpha({Q}))_f}.
\label{eqn:alfamu}
\end{equation}
Here a subscript ``lowest" is used to denote the lowest order $\alpha$.} This replacement can be taken as an effective renormalization of $\alpha$ at momentum transfer {$Q^2$}.
More details about this part can be found in Refs.~\cite{Denner:1991kt,Denner:2019vbn}. In our work, we set the relevant scale {$Q$} to $m_Z$ and resum over the contribution from light fermions, namely the running coupling constant is defined as
\begin{equation}
\alpha(m_Z)=\dfrac{\alpha(0)}{1-(\Delta\alpha(m_Z))_{f\neq\mathrm{top}}},
\label{eqn:alfamz}
\end{equation}
Here the label ``$f\neq\mathrm{top}$" denotes the contributions from light fermions, which {are} just the sum of $(\Delta \alpha)_{\mathrm{lepton}}$ and $(\Delta \alpha)_{\mathrm{q}}^{(5)}$ in Eq.~(\ref{eqn:ddalpha}).
This is how our $\alpha(m_Z)$ scheme is introduced. 
And it is regarded as  an ``on-shell'' definition of the running coupling constant at the scale $m_Z$ following the notation in PDG~\cite{Tanabashi:2018oca}.
Corresponding renormalization constants can be derived easily with Eq.~(\ref{eqn:alfamz}), which is
\begin{equation}
\delta Z_e(m_Z) = \delta Z_e(0) - \dfrac{1}{2}(\Delta\alpha(m_Z))_{f\neq\mathrm{top}}. 
\end{equation}

The above discussion is based on our choice of physical parametrization, $\{\alpha, m_W, m_Z\}$. Sometimes people use the Fermi constant $G_F$ as an input. Here we try to discuss the uncertainties arising from different input schemes.
The relationship between $G_F$ and above parameters is already well known, which is
\begin{equation}
G_F=\dfrac{\pi\alpha(1+\Delta r) }{\sqrt{2}m_W^2\left(1-\dfrac{m_W^2}{m_Z^2}\right)} .
\label{eqn:gf}
\end{equation}
$\Delta r$ can be calculated perturbatively. It vanishes at LO and its expression at NLO can be found in Ref.~\cite{Denner:1991kt}. 
Here we take $G_F$ as an observable, which is used to extract the value of certain renormalized physical parameter(s). 
The renormalization scheme in this case is exactly the same as the $\alpha(0)$ scheme.

If the input scheme is $\{\alpha, G_F, m_Z\}$, then this means the mass of $W$ boson has to be extracted with Eq.~(\ref{eqn:gf}).
However, there are some difficulties in this extraction:
\begin{enumerate}
\item There are new physics contributions in $\Delta r$, which make it vary in the IDM parameter space.
\item $\Delta r$ itself also depends on $m_W$.
\end{enumerate}
Hence, we use its SM value $\Delta r=0.03652$~\cite{Tanabashi:2018oca} as an approximation to estimate $m_W$, which leads to
\begin{equation}
m_W=m_Z\sqrt{\dfrac{1}{2}+\sqrt{\dfrac{1}{4}-\dfrac{\pi\alpha(1+\Delta r)}{\sqrt{2}G_Fm_Z^2}}}\approx 80.382~\mathrm{GeV}.
\end{equation}
It is very close to the value of $m_W$ we use in the $\{\alpha,m_W,m_Z\}$ scheme, which means the four inputs are precisely consistent. This indicates that the difference between these two schemes is negligible.

If the input scheme is $\{G_F, m_W, m_Z\}$, then the situation is similar,  since we have verified the consistency of the four inputs. However, in this case, similar to what we have done in Eq.~(\ref{eqn:alfamu}), we can introduce another effective coupling constant, which is 
\begin{equation}
\alpha_{G_F}=\dfrac{\alpha}{1-\Delta r}\approx \alpha(1+\Delta r).
\end{equation}
The value of $\alpha_{G_F}$ can be obtained directly from Eq.~(\ref{eqn:gf}), with no more approximation. We replace all the $\alpha$ with this $\alpha_{G_F}$ in our calculation. This can also be taken as an effective renormalization of $\alpha$, with the renormalzation constant
\begin{equation}
\delta Z_e|_{G_F}=\delta Z_e(0) - \dfrac{1}{2}\Delta r.
\end{equation}
Since $\Delta r$ also contains $\Pi(0)$, it will cancel with the one in $\delta Z_e(0)$.
Hence $\delta Z_e|_{G_F}$ no longer contains $\Pi(0)$, and the large logarithms discussed above are also resummed in this $\alpha_{G_F}$ scheme.

\begin{table}[!htb]
\centering
\begin{tabular}{|c|c|c|c|c|c|}
\hline
Case & 1 & 2 & 3 & 4 & 5  \\
\hline
Input & $\{\alpha,m_W,m_Z\}$ & $\{\alpha,m_W,m_Z\}$  & $\{\alpha, G_F, m_Z\}$  & $\{G_F, m_W, m_Z\}$ & $\{G_F, m_W, m_Z\}$ \\
\hline
Renormalization& $\alpha(0)$ & $\alpha(m_Z)$  & $\alpha(0)$  & $\alpha(0)$ &$\alpha_{G_F}$ \\
\hline
\end{tabular}
\caption{Several cases with different input and renormalization schemes are defined. }
\label{tab:schemes}
\end{table}
We list all the cases discussed above in Table~\ref{tab:schemes}. Cases 3 and 4 are very similar to case 1 due to the precise consistency among the four inputs. Hence, we skip them and consider the others. We choose a certain point in the IDM parameter space and calculate the cross section in  three renormalization schemes. The results are shown in Table~\ref{tab:scheme_comp}. 
It can be observed that  scheme dependence is greatly reduced at the NLO. 
However, it should be pointed out that the difference in the relative corrections is quite large, which originates from the difference in the LO results. Hence, when focusing on relative corrections, the difference in the LO results should be considered. And this can be easily estimated by the values of $\alpha$ in the table. For example, the relative correction in $\alpha(m_Z)$ scheme can be converted to $\alpha(0)$ ($\alpha_{G_F}$) scheme approximately by adding an extra piece of 12.9\% (5.1\%). 
After this, it can be seen that the difference in the remaining parts is around $1\%$, and is consistent with the difference in the NLO cross sections. 

\begin{table}[!htb]
\renewcommand\arraystretch{1.1}
\centering
\begin{tabular}{|c|c|c|c|c|c|c|c|c|}
\hline\hline
Case &Scheme& $\alpha^{-1}$ & $\sigma^0$ & $\sigma^1_{\mathrm{QED}}$ & $\sigma^1_{\mathrm{WEAK}}$ & $\sigma^1$&$\sigma^{\mathrm{NLO}}$  & $\Delta (\%)$\\
\hline
1&$\alpha(0)$       &137.036&23.678&0.468&~1.528&~1.996&25.674 & ~8.43\\\hline
2&$\alpha(m_Z)$  &128.943&26.744&0.562&-1.523 &-0.961&25.783 &-3.59\\\hline
5&$\alpha_{G_F}$&132.184&25.448&0.522&~0.128&~0.650&26.098 &~2.55\\
\hline\hline
\end{tabular}
\caption{Cross sections and relative corrections in different renormalization schemes are presented. The unit of cross sections is fb. IDM parameters are chosen as $m_{H}=m_{A}=m_{H^\pm}=200$ GeV, $\lambda_2=2$, and $\mu_2^2=0$. Values of the four inputs are taken as $\alpha(0)=1/137.036$, $m_W$=80.379 GeV, $m_Z$=91.1876 GeV, and $G_F=1.1663787\times 10^{-5}$ GeV$^{-2}$.}
\label{tab:scheme_comp}
\end{table}

\section{Wave function renormalization }
\label{sec:zh}

At the end of this section, we provide the explicit expression for $\delta Z_{H^\pm}$,  the only needed renormalization constant from the inert section in this work. It can be separated into four parts according to the boson inside the self-energy, namely
\begin{equation}
\delta Z_{H^\pm}=\delta Z_{H^\pm}^{W/G}+\delta Z_{H^\pm}^{Z/G}+\delta Z_{H^\pm}^\gamma+\delta Z_{H^\pm}^h, 
\end{equation}
with
\begin{equation}
\begin{aligned}
\delta Z_{H^\pm}^{W/G} =\frac{\alpha}{4\pi}\times &\frac{1}{4 s_W^2}\{
     ( B_0[m_{H^\pm}^2, m_{A^0}^2, m_{W}^2] + B_0[ m_{H^\pm}^2, m_{H^0}^2, m_{W}^2] ) \\&
 - 2( B_1[m_{H^\pm}^2, m_{A^0}^2, m_{W}^2] + B_1[ m_{H^\pm}^2, m_{H^0}^2, m_{W}^2] ) \\&
   -2m_{H^\pm}^2 (B_1'[ m_{H^\pm}^2, m_{A^0}^2, m_{W}^2]   +B_1'[ m_{H^\pm}^2, m_{H^0}^2, m_{W}^2]) \\&
   + [(m_{A^0}^2 + m_{H^\pm}^2) - {(m_{A^0}^2 - m_{H^\pm}^2)^2}/{m_W^2}]
 B_0'[m_{H^\pm}^2, m_{A^0}^2, m_{W}^2]  \\&
 + [(m_{H^0}^2 + m_{H^\pm}^2) - {(m_{H^0}^2 - m_{H^\pm}^2)^2}/{m_W^2}]
 B_0'[m_{H^\pm}^2, m_{H^0}^2, m_{W}^2]\} \\
\delta Z_{H^\pm}^{Z/G} = \frac{\alpha}{4\pi} \times  &g_H^2\{
 B_0[m_{H^\pm}^2, m_{H^\pm}^2, m_{Z}^2] 
 - 2B_1[ m_{H^\pm}^2, m_{H^\pm}^2, m_{Z}^2] 
 \\&
 +2m_{H^\pm}^2 B_0'[ m_{H^\pm}^2, m_{H^\pm}^2, m_{Z}^2]
 -2m_{H^\pm}^2  B_1'[ m_{H^\pm}^2, m_{H^\pm}^2, m_Z^2] \} \\
\delta Z_{H^\pm}^\gamma =\frac{\alpha}{4\pi}\times &\{
 B_0[m_{H^\pm}^2, 0, m_{H^\pm}^2]  
 -2 B_1[m_{H^\pm}^2, m_{H^\pm}^2, 0]  
 \\&
+2 m_{H^\pm}^2 B_0'[m_{H^\pm}^2, 0, m_{H^\pm}^2]   
 -2 m_{H^\pm}^2B_1'[m_{H^\pm}^2, m_{H^\pm}^2, 0]
 \}\\
\delta Z_{H^\pm}^h = \frac{\alpha}{4\pi}\times &
 \frac{-(m_{H^\pm}^2-\mu_2^2)^2}{m_W^2 s_W^2}
 B_0'[m_{H^\pm}^2, m_{h^0}^2, m_{H^\pm}^2] 
 \end{aligned}
 \end{equation}
and 
\begin{equation}
B_i'[m_{H^\pm}^2,m_1^2,m_2^2] \equiv \dfrac{\partial B_i[k^2,m_1^2,m_2^2]}{\partial k^2} \biggr|_{k^2=m_{H^\pm}^2} .
\end{equation}
Here the two-point functions $B_0$ and $B_1$ are defined as 
\begin{eqnarray}
	B_0[p_1^2,m_1^2, m_2^2]&=&\frac{(2 \pi {\mu_r})^{4-D}}{i \pi^2 } \int d^Dq \, \frac{1}{(q^2-m_1^2)\big((q+p_1)^2-m_2^2\big)}  \nonumber \\
	B_1[p_1^2,m_1^2, m_2^2]&=& \frac{1}{2}\Big\{
        A_0[m_1^2] - A_0[m_2^2] - \left[p_1^2 + m_1^2 - m_2^2\right] B_0[p_1^2, m_1^2, m_2^2] \Big\}
\end{eqnarray}
with $\mu_r$ being the renormalization scale and $A_0$ being the one-point function given by
\begin{eqnarray}
    A_0(m_1^2) &=& \frac{(2 \pi {\mu_r})^{4-D}}{i \pi^2}\int_{}^{}{d^Dq \frac{1}{q^2-m_1^2}}.
\end{eqnarray}
{In our work, as shown in Eq.~(\ref{eqn:e0}), the mass parameter $\mu$ in bare charge is replaced by $\mu_r$ after the renormalization. Consequently,  $\mu_r$ appears instead of $\mu$ in above one- and two-point functions.
}

\section{More details about treatment on IR divergences}
\label{sec:IR}
As mentioned in the main text, IR  divergences in this work are regularized with a small fictitious photon mass $\lambda$. 
Meanwhile, two cutoffs, $\Delta E$ and $\Delta\theta$, are introduced to deal with the IR singularities in real correction processes based on the two cutoff phase space slicing method\cite{Harris:2001sx}. The three-body phase space of the real correction process $e^+e^-\rightarrow H^+H^-\gamma$ is then divided into three parts:
\begin{itemize}
\item Soft ($S$) part: Where the energy of photon $E_\gamma$ is smaller than $\Delta E$.
\item Hard collinear ($HC$) part: Where $E_\gamma\ge\Delta E$ and the angle between photon and the beam $\theta_\gamma$ is smaller than $\Delta\theta$.
\item Hard noncollinear ($H\overline{C}$) part: The remaining, which is finite.
\end{itemize}
The full NLO corrections are then separated into four parts, as given in Eq.~(\ref{fullcor}):
\begin{equation}
d\sigma^{1}=d\sigma_V(\lambda)+d\sigma_S(\lambda,\Delta E)+d\sigma_{HC+CT}(\Delta E,\Delta\theta)+d\sigma_{H\overline{C}}(\Delta E,\Delta\theta) 
\label{fullcor}
\end{equation}
Here $d \sigma_V$ denotes the virtual correction, including loop diagrams and counter terms from renormalization.
And $CT$ in the third term of rhs denotes the extra contribution arising from the structure function of the incoming electron and positron.

In our work, the first two parts, $d \sigma_V$ and $d \sigma_S$, are obtained using \texttt{FeynArts} and \texttt{FormCalc}~\cite{Hahn:2000kx,Hahn:1998yk,Hahn:2006qw} packages, in which numerical evaluations of the scalar integrals are done with
\texttt{LoopTools}~\cite{Hahn:1999mt,Hahn:2010zi}. The other two parts, $d\sigma_{HC+CT}$ and $d\sigma_{H\overline{C}}$, are obtained with the help of \texttt{FDC}~\cite{Wang:2004du} and \texttt{BASES}~\cite{Kawabata:1995th}.

The soft part is calculated in the soft limit.  After the phase space integration of the soft photon, it can be expressed as the product of a factor and the LO result. The result of this part can be found in the literature~\cite{Arhrib:1998gr,Guasch:2001hk}.
 For completeness, we give the analytical expressions here:
\begin{equation}
\begin{aligned}
d\sigma_S=&-\frac{\alpha}{\pi}d\sigma^0\times\left\{
4 \log \frac{2 \Delta E}{\lambda}-2 \log \frac{2 \Delta E}{\lambda} \log \frac{s}{m_{e}^{2}}
+\log \frac{m_{e}^{2}}{s}+\frac{\pi^{2}}{3}+\frac{1}{2} \log ^{2} \frac{m_{e}^{2}}{s}   \right.\\ 
&+\frac{1+\beta^{2}}{\beta} \log \frac{2 \Delta E}{\lambda} \log \left(\frac{1-\beta}{1+\beta}\right) 
+\frac{1}{\beta} \log \left(\frac{1-\beta}{1+\beta}\right)
+\frac{1+\beta^{2}}{\beta}\left[\operatorname{Li}_{2}\left(\frac{2 \beta}{1+\beta}\right)+\log ^{2}\left(\frac{1-\beta}{1+\beta}\right)\right] \\ 
&+4 \log \frac{2 \Delta E}{\lambda} \log \frac{m_{H^{\pm}}^{2}-u}{m_{H^{\pm}}^{2}-t}
+2\left[\operatorname{Li}_{2}\left(1-\frac{s(1-\beta)}{2\left(m_{H^{\pm}}^{2}-t\right)}\right)+\operatorname{Li}_{2}\left(1-\frac{s(1+\beta)}{2\left(m_{H^{\pm}}^{2}-t\right)}\right)\right.\\ 
&\left.\left.-\operatorname{Li}_{2}\left(1-\frac{s(1-\beta)}{2\left(m_{H^{\pm}}^{2}-u\right)}\right)-\operatorname{Li}_{2}\left(1-\frac{s(1+\beta)}{2\left(m_{H^{\pm}}^{2}-u\right)}\right)\right]\right\} 
\end{aligned}
\end{equation}
where Li$_2$  represents the known dilogarithm function. 
\texttt{FormCalc} has the option to include soft bremsstrahlung automatically when calculating the virtual corrections. We have confirmed that the results from \texttt{FormCalc} are consistent with the above expression. Meanwhile, by varying the value of $\lambda$ inside \texttt{FormCalc}, we confirm the $\lambda$ independence of our results.

The hard noncollinear part, $d\sigma_{H\overline{C}}$, needs no more special treatment. It is obtained using traditional Monte Carlo integration techniques with the \texttt{FDC} package, in which \texttt{BASES}~\cite{Kawabata:1995th} is used for multidimensional numerical integration.

The hard collinear part is obtained in the collinear limit as
\begin{equation}
\begin{aligned}
d\sigma_{HC} =&\dfrac{\alpha}{2\pi}\left[\dfrac{1+z^2}{1-z}\log\dfrac{\Delta\theta^2+4m_e^2/s}{4m_e^2/s}-\dfrac{2z}{1-z}\dfrac{\Delta\theta^2}{\Delta\theta^2+4m_e^2/s}\right]d\sigma^0(zp_1)dz +(p_1\Leftrightarrow p_2)   \label{eqn:HC} 
\end{aligned}
\end{equation}
Here $z$ denotes the energy fraction of the electron (positron) after the emission of collinear  a photon; hence, $0\leq z \leq 1-\delta_s$. 
We have used a dimensionless parameter $\delta_s=2\Delta E/\sqrt{s}$ to replace $\Delta E$.
We have used $d\sigma^0\equiv d\sigma^0(p_1,p_2)$, $d\sigma^0(zp_1)\equiv d\sigma^0(zp_1,p_2)$, and $d\sigma^0(zp_2)\equiv d\sigma^0(p_1,zp_2)$ for convenience. 

{ {It is found that in the limit $\Delta\theta^2\gg m_e^2/s$, the hard collinear part given in Eq. (\ref{eqn:HC}) can be expressed in a simpler form as given below
\begin{equation}
\begin{aligned}
d\sigma_{HC} \xrightarrow{\Delta\theta^2\gg m_e^2/s}&\dfrac{\alpha}{2\pi}\left[\dfrac{1+z^2}{1-z}\log\dfrac{\Delta\theta^2s}{4m_e^2}-\dfrac{2z}{1-z}\right]\times\biggl[d\sigma^0(zp_1)+d\sigma^0(zp_2)\biggr]dz .
\label{eqn:HCB}
\end{aligned}
\end{equation}
We will use the form given in Eq. (\ref{eqn:HCB}) as our default formula to compute the full cross section given in Eq. (\ref{fullcor}). It is interesting to explore the difference between the results of Eq. (\ref{eqn:HC}) and Eq. (\ref{eqn:HCB}), and we will explore this issue near the end of this section.}}

One-loop radiation correction includes collinear singularities, which lead to terms proportional to $\log(m_e)$. 
Some of them are canceled when summing up virtual and real corrections. 
Some of them are absorbed into the redefinition of the running coupling constant, as shown in Eq.~(\ref{eqn:alfamz}).
But there are still some remaining. 
To deal with this, the structure function approach~\cite{Kuraev:1985hb} is applied.
According to the approach, the cross section of $e^+e^-$ annihilation can be expressed as 
\begin{equation}
d\overline{\sigma}_{e^+e^-}(p_1,p_2)=\sum_{ij}dx_1 dx_2 f_{ie^+}(x_1,\overline{Q}^2) f_{je^-}(x_2,\overline{Q}^2)d\sigma_{ij}(x_1p_1,x_2p_2)
\label{eqn:factorization}
\end{equation}
Here $f_{ia}(x,\overline{Q}^2)$ is the so-called structure function, which denotes the possibility to find parton $i$ with energy faction $x$ from particle $a$ at scale $\overline{Q}$.  In Eq.~(\ref{eqn:factorization}), all remaining collinear singularities are absorbed into the structure functions and resummed there. The parton-level cross section $d\sigma_{ij}$ is then free of those large logarithms. The particle-level cross section $d\overline{\sigma}_{e^+e^-}$ is obtained by convoluting the partonic cross section with the structure functions. 
This factorization theorem is extended from QCD with $i,a\in \{e^+,e^-,\gamma\}$. 
Strictly speaking, what we are studying in this work is the parton-level cross section of $H^+H^-$ production only. Since we are interested in the NLO corrections in the whole parameter space, this is sufficient. But if we want to study the exact cross section of production, convolution with the structure functions is needed.

In QED, all the structure functions  $f_{ia}(x,\overline{Q}^2)$ are perturbatively calculable. They can be expanded into a series of $\alpha$, namely 
\begin{equation}
f_{ia}(x,\overline{Q}^2)=f_{ia}^0(x,\overline{Q}^2)+f_{ia}^1(x,\overline{Q}^2)+{\cal O}(\alpha^2),
\end{equation}
Meanwhile it is obvious that the first term in above expansion always takes the form
\begin{equation}
f_{ia}^0(x,\overline{Q}^2)=\delta(1-x)\delta_{ia}.
\label{eqn:sf0}
\end{equation}
Based on this, in this work up to one-loop level, Eq.~(\ref{eqn:factorization}) can be rewritten into (using $f_{ee}\equiv f_{e^-e^-}=f_{e^+e^+}$)
\begin{equation}
d\overline{\sigma}_{e^+e^-}(p_1,p_2)=dx_1 dx_2 f_{ee}(x_1,\overline{Q}^2) f_{ee}(x_2,\overline{Q}^2)d\sigma_{e^+e^-}(x_1p_1,x_2p_2)
\label{eqn:factorization1}
\end{equation}
Expanding both sides of Eq.~(\ref{eqn:factorization1}) to NLO and applying Eq.~(\ref{eqn:sf0}), it is easy to find 
\begin{equation}
\begin{aligned}
d\overline{\sigma}_{e^+e^-}^0(p_1,p_2)=&d\sigma_{e^+e^-}^0(p_1,p_2) \\
d\overline{\sigma}_{e^+e^-}^1(p_1,p_2)=&d\sigma_{e^+e^-}^1(p_1,p_2) +dx_1 f_{ee}^1(x_1,\overline{Q}^2) d\sigma_{e^+e^-}^0(x_1p_1,p_2) \\
&+dx_2 f_{ee}^1(x_2,\overline{Q}^2)d\sigma_{e^+e^-}^0(p_1,x_2p_2),
\end{aligned}
\end{equation}
which leads to 
\begin{equation}
d\sigma_{e^+e^-}^1(p_1,p_2)=d\overline{\sigma}_{e^+e^-}^1(p_1,p_2)-dz f_{ee}^1(z,\overline{Q}^2)\biggl[d\sigma_{e^+e^-}^0(zp_1,p_2)+d\sigma_{e^+e^-}^0(p_1,zp_2)\biggr]
\label{eqn:CT0}
\end{equation}
The latter term on the rhs is just the counterterm mentioned  above.

On the other hand, $f_{ee}$ satisfies following equation~\cite{Kuraev:1985hb}:
\begin{equation}
f_{ee}(x,\overline{Q}^2)=\delta(1-x)+\int\limits_{m_e^2}^{\overline{Q}^2}\dfrac{\alpha(\overline{Q}_1^2)}{2\pi}\dfrac{d\overline{Q}_1^2}{\overline{Q}_1^2}\int\limits_x^1\dfrac{dz}{z}P_{ee}^+(z)f_{ee}\left(\dfrac{x}{z},\overline{Q}_1^2\right),
\end{equation}
with
\begin{equation}
P_{ee}^+(z)=\dfrac{1+z^2}{(1-z)_+}+\dfrac{3}{2}\delta(1-z),
\label{eqn:pee}
\end{equation}
being the regularized Altarelli-Parisi  splitting function. This equation can be solved recursively according to the power of $\alpha$, which leads to 
\begin{equation}
f_{ee}^1(x,\overline{Q}^2)=\dfrac{\alpha}{2\pi}\log\dfrac{\overline{Q}^2}{m_e^2}P_{ee}^+(x).
\label{eqn:ff1}
\end{equation}
In this work, the scale $\overline{Q}$ is taken as the energy of beam in the center-of-mass frame, $\overline{Q}=\sqrt{s}/2$. 
The CT parts are finally obtained via Eqs.~(\ref{eqn:CT0}) and (\ref{eqn:ff1}) as [after same notations as in Eq.~(\ref{eqn:HC}) are taken]
\begin{equation}
d\sigma_{CT} =-\dfrac{\alpha}{2\pi}\log\dfrac{s}{4m_e^2}P_{ee}^+(z)\biggl[d\sigma^0(zp_1)+d\sigma^0(zp_2)\biggr]dz
\end{equation}
with $0\leq z\leq 1$.

The combination of $d\sigma_{HC}$ and $d\sigma_{CT}$ can be separated into two parts according to the range of $z$, $d\sigma_{HC+CT}=d\sigma_{HC+CT}^{*}+d\sigma_{SC}$.
In the range $0\leq z \leq 1-\delta_s$, it gives 
\begin{equation}
d\sigma_{HC+CT}^{*} =\dfrac{\alpha}{2\pi}\left[\dfrac{1+z^2}{1-z}\log\Delta\theta^2-\dfrac{2z}{1-z}\right]\times\biggl[d\sigma^0(zp_1)+d\sigma^0(zp_2)\biggr]dz,
\end{equation}
while in the range $1-\delta_s \leq z \leq 1$, it gives (only $d\sigma_{CT}$ contributes to this part)
\begin{equation}
d\sigma_{SC} =-\dfrac{\alpha}{\pi}\log\dfrac{s}{4m_e^2} \left[\dfrac{3}{2}+2\log\delta_s\right]d\sigma^0.
\end{equation}
Now all the parts in Eq.~(\ref{fullcor}) are available.

\begin{table}[!htb]
\centering
\begin{tabular}{|c|c|c|c|c|c|c|c|}
\hline\hline
$\delta_s$& $\sigma_{V+S}$ & $\sigma_{HC+CT}^*$& $\sigma_{SC}$ &$\sigma_{H\overline{C}}$&SUM\\
\hline
$10^{-1}$     &~-6.364(0)&~-1.047(0)&~5.371(0)&~1.118(0)&-0.922(0)\\\hline
$10^{-2}$     &-14.621(0)&~-5.074(1)&13.338(1)&~5.399(0)&-0.958(1)\\\hline
$10^{-3}$     &-22.878(0)&~-9.531(1)&21.304(1)&10.144(1)&-0.961(2)\\\hline
$10^{-4}$     &-31.135(0)&-14.031(2)&29.270(2)&14.935(1)&-0.961(3)\\\hline
$10^{-5}$     &-39.392(0)&-18.536(3)&37.237(2)&19.730(2)&-0.961(4)\\\hline
\hline
$k$& $\sigma_{V+S}$ & $\sigma_{HC+CT}^*$& $\sigma_{SC}$ &$\sigma_{H\overline{C}}$&SUM\\
\hline
$2^{-6}$    &-29.641(0)&-9.531(1)&28.067(2)&10.144(1)&-0.961(2)\\\hline
$2^{-4}$    &-27.387(0)&-9.531(1)&25.813(1)&10.144(1)&-0.961(2)\\\hline
$2^{-2}$    &-25.132(0)&-9.531(1)&23.558(1)&10.144(1)&-0.961(2)\\\hline
$2^{-1}$    &-24.005(0)&-9.531(1)&22.431(1)&10.144(1)&-0.961(2)\\\hline
$2^{0}$     &-22.878(0)&-9.531(1)&21.304(1)&10.144(1)&-0.961(2)\\\hline
$2^{1}$     &-21.751(0)&-9.531(1)&20.177(1)&10.144(1)&-0.961(2)\\\hline
$2^{2}$     &-20.624(0)&-9.531(1)&19.050(1)&10.144(1)&-0.961(2)\\\hline
$2^{4}$     &-18.370(0)&-9.531(1)&16.795(1)&10.144(1)&-0.962(2)\\\hline
$2^{6}$     &-16.115(0)&-9.531(1)&14.541(1)&10.144(1)&-0.961(2)\\\hline
\end{tabular}
\caption{\label{tab:check1}Data used in the checks on the independence of $\delta_s$ and $m_e$ in unit of fb are shown. The third column is obtained by using  Eq. (\ref{eqn:HCB}).
Numbers in the brackets are integration errors to the last digits of the data.
} 
\end{table}
\begin{table}
\centering
\begin{tabular}{|c|c|c|c|c|c||c|c|}
\hline
$\Delta\theta$& $\sigma_{V+S}$ & $\sigma_{HC+CT}^*$& $\sigma_{SC}$ &$\sigma_{H\overline{C}}$&SUM&$\sigma_{HC+CT}^{*,\mathrm{origin.}}$&$\mathrm{SUM}^{\mathrm{origin.}}$\\
\hline
$10^{-1}$     &-22.878(0)&~-3.605(0)&21.304(1)&~4.214(0)&-0.965(1) &~-3.605(0)&-0.965(1)\\\hline
$10^{-2}$     &-22.878(0)&~-6.568(1)&21.304(1)&~7.180(1)&-0.962(2) &~-6.568(1)&-0.962(2)\\\hline
$10^{-3}$     &-22.878(0)&~-9.531(1)&21.304(1)&10.144(1)&-0.961(2) &~-9.531(1)&-0.961(2)\\\hline
$10^{-4}$     &-22.878(0)&-12.495(2)&21.304(1)&13.107(1)&-0.962(2) &-12.494(2)&-0.961(2)\\\hline
$10^{-5}$     &-22.878(0)&-15.458(2)&21.304(1)&16.019(1)&-1.013(2) &-15.406(2)&-0.961(2)\\\hline
$10^{-6}$     &-22.878(0)&-18.421(2)&21.304(1)&17.459(2)&-2.536(3) &-16.845(2)&-0.960(3)\\\hline
\hline
\end{tabular}
\caption{\label{tab:check2}Data used in the checks on the independence of $\Delta\theta$ in unit of fb are shown. The seventh column and the eighth column with the superscript ``origin." denote the results by using Eq. (\ref{eqn:HC}). In contrast, the third column and the sixth column denote the results by using  Eq. (\ref{eqn:HCB}). Numbers in the brackets are integration errors to the last digits of the data.} 
\end{table}

\begin{figure}[!htb]
\begin{center}
\includegraphics[width=0.31\textwidth]{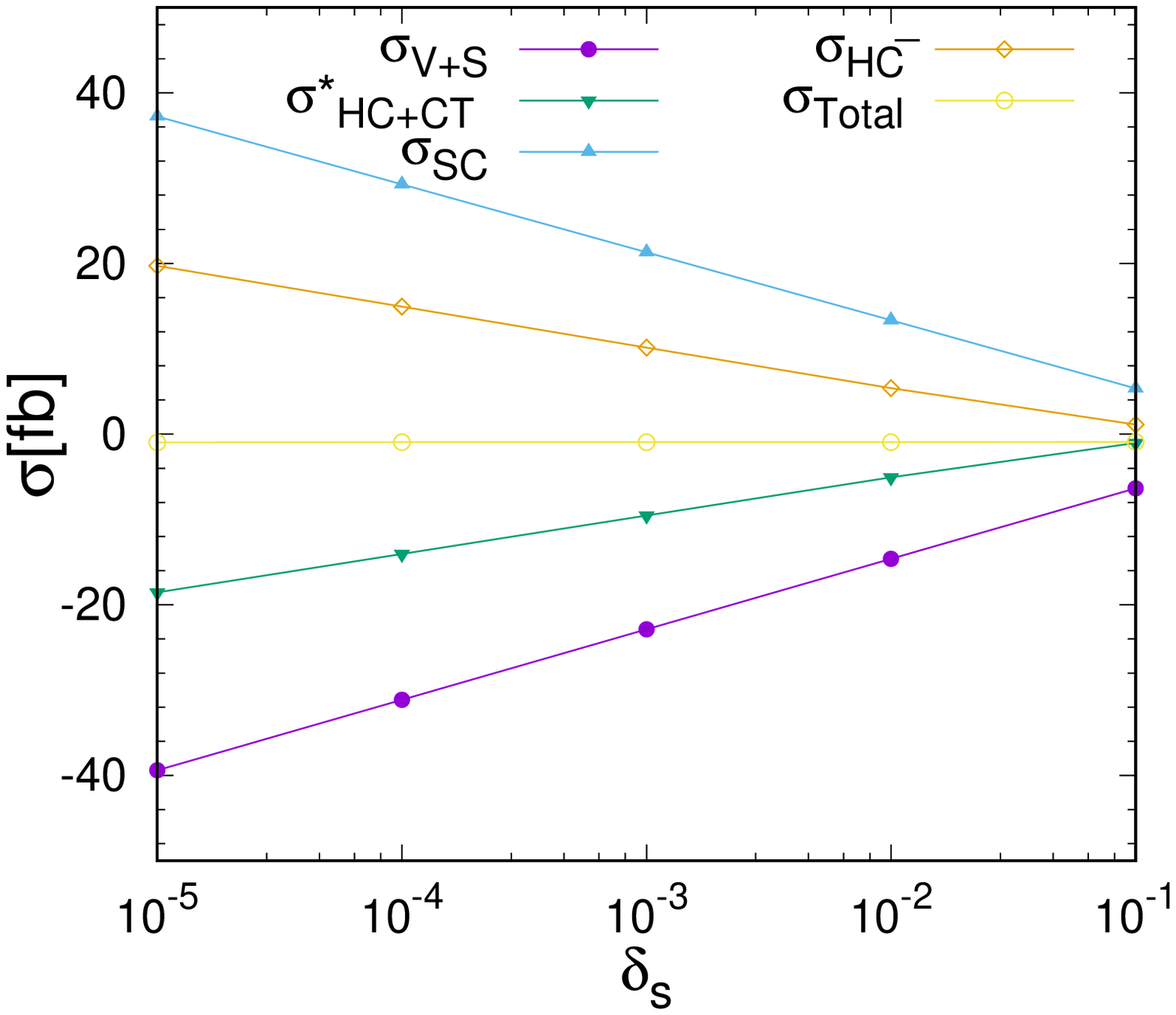}
\includegraphics[width=0.31\textwidth]{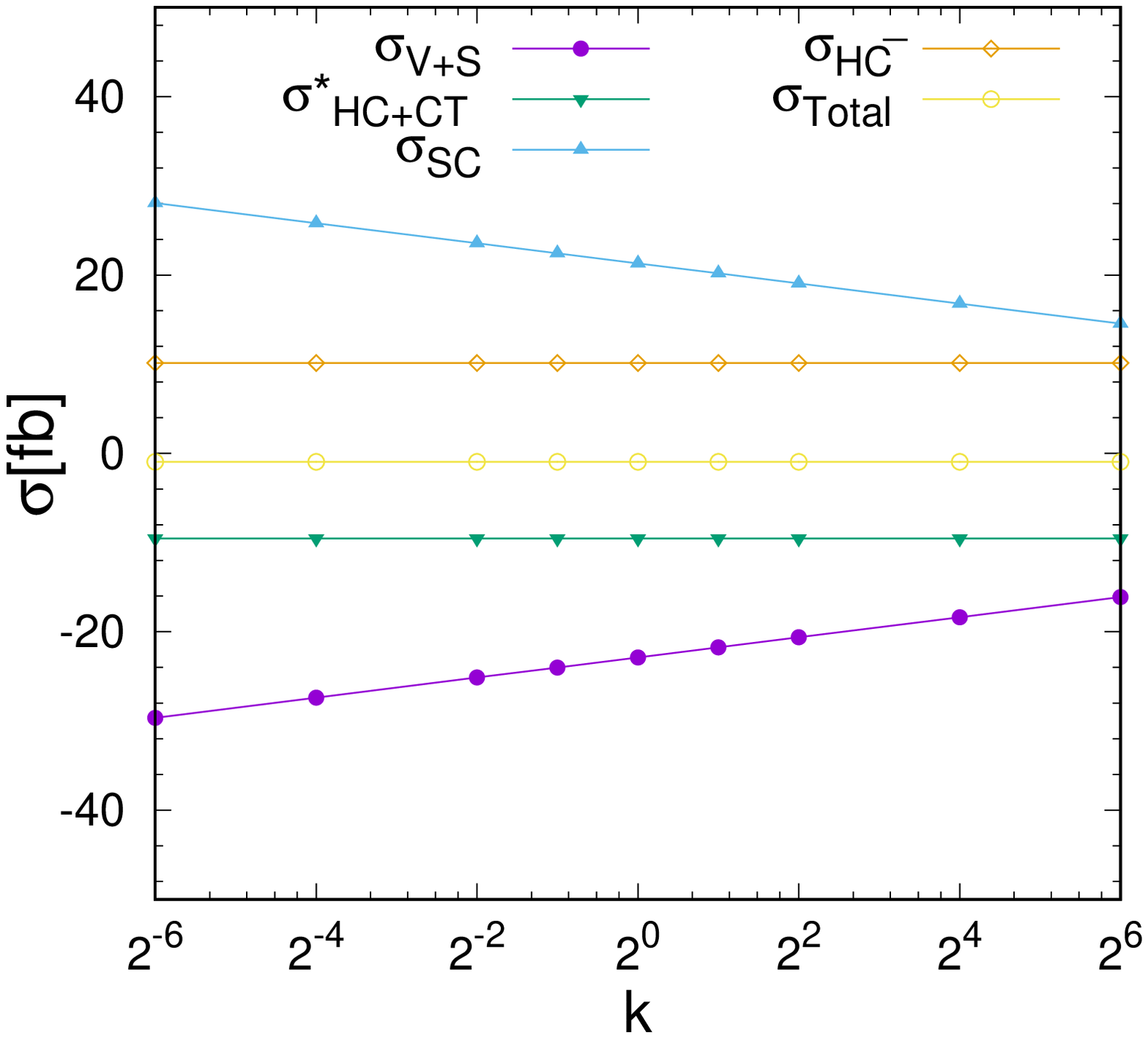}
\includegraphics[width=0.31\textwidth]{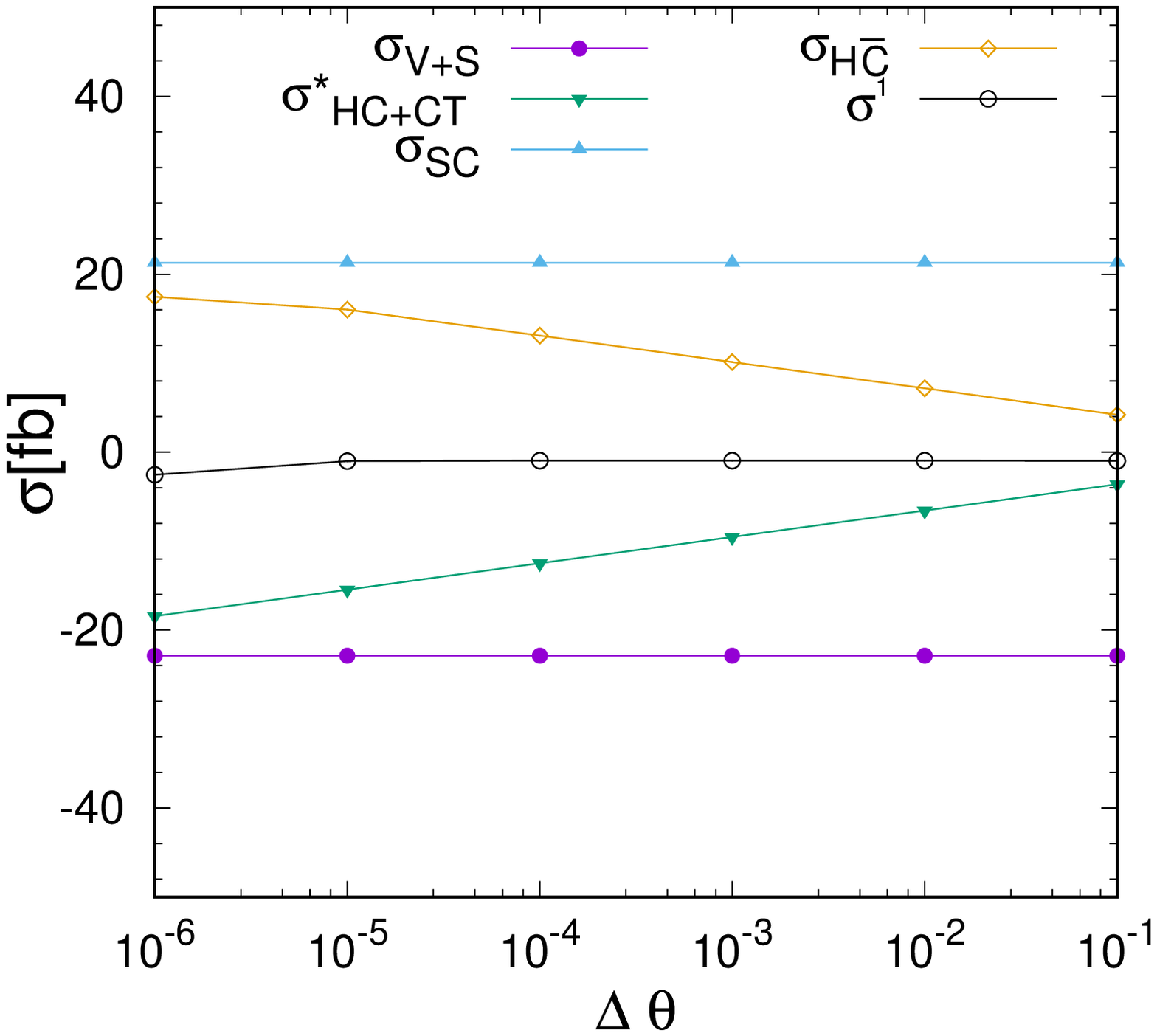}
\caption{One-loop corrections of $e^+e^-\rightarrow H^+H^-$ as functions of $\delta_s$, $k$, and $\Delta\theta$ are demonstrated.}
\label{figcheck}
\end{center}
\end{figure}

It should be stressed that $\Delta E$ ($\delta_s$) and $\Delta\theta$ are unphysical cutoffs we introduced to deal with IR and collinear singularities. Our final results should not depend on them. In order to check this, we choose a certain point (same as the one used in Table \ref{tab:scheme_comp}) in the IDM parameter space and compare the NLO corrections $\sigma^1$ obtained with different cutoffs. 
{
The results are shown in Fig.~\ref{figcheck}, with corresponding data in Tables ~\ref{tab:check1} and \ref{tab:check2}.

From the first subfigure, it can be seen that the independence on $\delta_s$ is found in a wide range and $\delta_s=10^{-3}$ is used as our default choice. }

Collinear divergences in our calculation appear as terms proportional to $\log(m_e)$. After including the counterterm from the structure function of an electron, such divergent terms should vanish in the final result. In order to check this, we vary the mass of the electron by a factor of $k$ from $2^{-6}$ to $2^6$, namely $m_e$ is taken $k\times 0.511$ MeV. The cancellation is shown in the second subfigure of Fig.~\ref{figcheck}, from which we can see that the result remains unchanged as $k$ varies. {Furthermore, singular terms appear only in the $\sigma_{V+S}$ and $\sigma_{SC}$ parts.}

{{In the third subfigure, we can see that the result becomes cut dependent when $\Delta\theta$ is smaller than $10^{-4}$. 
It can be attributed to the fact that Eq.~(\ref{eqn:HCB}) can only hold when $\Delta\theta\gg m_e/\sqrt{s}\sim 2\times10^{-6}$. 

It is interesting to compare the results computed by using Eqs. (\ref{eqn:HCB}) and  (\ref{eqn:HC}), which are provided in Table~\ref{tab:check2}. It is clear that the result which is labeled by a superscript ``origin." is almost independent of  the values of $\Delta\theta$ (when $\Delta\theta$ is small enough), and the difference between the results calculated by using Eqs. (\ref{eqn:HCB}) and (\ref{eqn:HC})  is tiny when $\Delta\theta$ is chosen appropriately (say $\Delta \theta \sim 10^{-3}$ or larger). In our practical computation, we choose $\Delta\theta=10^{-3}$ in this work.
}}

{
In addition to the above checks for independence on $\lambda$, $\delta_s$, $\Delta\theta$, and $\log(m_e)$, our results also passed many other self-checks:
\begin{itemize}
\item The IDM model used in \texttt{FDC} is generated with its own code. We have confirmed that both \texttt{FDC} and \texttt{FormCalc} give the same results at LO.
\item The soft part $d\sigma_S$ has been calculated individually and checked with analytic results, as mentioned before.
\item The other parts of real emission, $d\sigma_{HC+CT}$ and $d\sigma_{H\overline{C}}$, obtained with FDC are also tree-level calculations. This part of FDC has been used in many other calculations (see e.g. \cite{Gong:2009kp, Gong:2012ug}).
\item \texttt{FormCalc} is a public package that has been used in many one-loop EW calculations (see e.g. \cite{Sun:2016bel,Gong:2016jys})
as well as in our  previous works~\cite{Xie:2018yiv,Abouabid:2020eik}.
\item We have also checked our model file using the output from \texttt{FeynRules}.
\end{itemize}
However, it should be noted that further cross-checks are still important and welcome.
}



\bibliographystyle{JHEP}
\bibliography{biblio}
\end{document}